# A reconstruction of quantum theory for nonspinning particles

Ulf Klein

Institute for Theoretical Physics, University of Linz, A-4040 Linz, Austria

**Within the framework of the individuality interpretation of quantum theory (QT), the basic equations of QT cannot be derived from the basic equations of classical mechanics (CM). The unbridgeable gap between CM and QT is given by the fact that a certain system which is described in CM by a finite number of degrees of freedom requires an infinite number in QT. The standard quantization method, which is conceptually closely linked to the individuality interpretation, is limited to finding structural similarities between observables and operators. The fundamental question *why* one must move from a finite number to an infinite number of degrees of freedom, remains unanswered. This gap can only be closed if probabilistic aspects are already taken into account in the classical area. This may be done by taking the uncertainty in initial conditions into account. In this probabilistic version of mechanics (PM), a system is mathematically described as an ensemble, with an infinite number of degrees of freedom, thus bridging the gap mentioned above. This step then enables the reconstruction of QT, in particular the derivation of the Schrödinger equation, from PM. This work is the third in a series of works in which this program is carried out. The method used here differs from the previous one and allows a better understanding of the structural differences between classical physics and QT. The derivation of the Schrödinger equation essentially takes place in two steps: a projection from phase space to configuration space and a linearization. Some contradictions of the individuality interpretation are analyzed and eliminated from the point of view of the ensemble interpretation.**

Ulf Klein: ulf.klein@jku.at

## 1 Introduction

The present work arose from unresolved questions related to the interpretation of quantum theory (QT). For this reason, we begin this work with a discussion of the different interpretations of QT.

An unresolved question of fundamental importance is which *object* is actually described by QT. The methodological principle of reductionism often plays a decisive role in answering this question, even if this is rarely explicitly stated. It is a fact that in the range of atomic dimensions and below, classical mechanics (CM) loses its validity and is replaced by QT. It seems reasonable to conclude from this that CM is a special case of QT in the sense of a reductionist hierarchy and that, in particular, both theories describe the same objects, namely individual particles. This is the widespread *individuality interpretation* of QT.

A single quantum mechanical particle is described - in the context of the individuality interpretation - by the single-particle Schrödinger equation. If we compare the concept of *state* in both theories, we come across irreconcilable differences. In CM, the state of a particle (at a fixed time t) is determined by a point in phase space with 6 components. In the transition to QT, this point in phase space becomes a point in a function space (Hilbert space) with an infinite number of components. The same enormous difference becomes visible when one considers the construction of the single-particle Schrödinger equation in the context of the individuality interpretation. In the simplest case, this *quantization process* looks like this: in the formula $E = H(q,p)$, where $H$ is the Hamiltonian function, the quantities $q, p, E$ are replaced by operators

$$\hat{q} = q, \quad \hat{p} = \frac{\hbar}{\imath}\frac{\mathrm{d}}{\mathrm{d}q}, \quad \hat{E} = -\frac{\hbar}{\imath}\frac{\mathrm{d}}{\mathrm{d}t} \quad , \tag{1}$$

and these are then allowed to act on a wave function $\psi$. This recipe works quite well. However, it is completely unclear why it works. One can ask two questions in particular:

1. Why is it necessary to introduce a function space, i.e. to introduce an infinite number of degrees of freedom to describe the state of a single particle?

2. Why do the operators have exactly this form and not another?

The individuality interpretation does not provide an answer to any of these questions. The first two of the rules (1) are used in the quantization method of the individuality interpretation to derive the form of arbitrary operators. In doing so, the explanation of question 1. is completely omitted and only an attempt is made to construct a structural similarity between the algebra of classical observables in phase space and the algebra of non-commutative operators in Hilbert space [21].

The individuality interpretation also leads to contradictions when analysing the results of experiments [16, 15, 2, 28, 40]. There are two classes of attempts to integrate the individuality interpretation into QT *despite* these contradictions. Theories belonging to the first class take it for granted that QT is incomplete, because it cannot explain the behavior of single particles, and conclude from this that it must be somehow modified, extended or completed. The most prominent proponent of this deterministic attitude was Albert Einstein. In his famous work with Podolsky and Rosen, Einstein showed that QT is predictively incomplete. Furthermore, due to his basic philosophical attitude, he assumed that QT had to be replaced by a more complete (deterministic) theory, i.e. that it was also incomplete in this metaphysical respect. These two concepts of (in)completeness are *completely* different. Theories belonging to the second class leave the structure of QT unchanged, but nevertheless assume that QT is able to describe single particles. An attempt is made to eliminate the contradictions that arise by means of suitable (very creative) interpretations.

The second possible answer to the basic question mentioned at the beginning is that QT only makes statements about statistical ensembles [15, 2]. The present work is based on this assumption. We try to better understand QT as it is and as it has been confirmed millions of times, leaving aside the above question of its metaphysical (in)completeness. (We do not doubt the metaphysical completeness of QT). In principle, this statistical or *ensemble interpretation* of QT makes no statements about the behavior of individual particles; however, this loss of supposed information is fortunately accompanied by the absence of contradictions associated with exactly this "information".

The infinite number of degrees of freedom in QT is associated with the occurrence of probabilities. The latter can be described in a natural way within the framework of an ensemble theory. Let us consider three different but related physical theories that differ with regard to the role of probability.

1. In nature, there are localized objects that have the property that the values of position and velocity at a certain time determine the values of these quantities at any later time. These point particles obey Newton's equations, which in turn are closely linked to the structure of (non-relativistic) space-time. Whenever "particles" are mentioned, these objects are meant. Instead of Newton's equations of motion, we use the canonical equations

$$\dot{q}_k = \frac{\partial H(q,p)}{\partial p_k}, \quad \dot{p}_k = -\frac{\partial H(q,p)}{\partial q_k}, \tag{2}$$

as basic equations of CM. For the purpose of illustration, we may think of a single particle without interaction, with coordinates $q_k$, conjugate momenta $p_k$, and Hamilton's function $H(q,p)$. Law (2) is of course deterministic in the sense that the state $(q(t),p(t))$ at any time $t$ is determined by the state $(q(0),p(0))$ at time $t=0$.

2. Let us now consider a situation in which the initial values $(q(0),p(0))$ are not exactly known but are subject to uncertainty. To describe such a situation, we have to consider a continuum of solutions of (2) and define an appropriate probability distribution $\rho_0(q,p)$ at time $t=0$. The time-dependence of the trajectories will still be determined by Eq. (2). The continuous set of particles trajectories constructed in this way must then be converted to the usual field-theoretical form. This process is called "Lagrangian to Eulerian transition" [3, 32]. Field variables are introduced whose temporal and spatial distribution in phase space is determined by the solutions of the equations of motion (2). This leads to the Liouville equation

$$\frac{\partial \rho}{\partial t} + \frac{\partial \rho}{\partial q_k}\frac{\partial H}{\partial p_k} - \frac{\partial \rho}{\partial p_k}\frac{\partial H}{\partial q_k} = 0. \tag{3}$$

for the probability density $\rho(q, p, t)$. We will refer to the theory defined by (3) as probabilistic mechanics (PM), as it differs from CM in essential points despite its close connection to (2). This concerns not only the mathematical structure but also the structure of the physical predictions and the nature of the states. Although PM is essentially a statistical theory, it naturally also contains a strong deterministic component. The term "cryptodeterminism" introduced by Moyal describes this aptly [45].

3. The third physical theory we consider is QT; it is again sufficient to consider only the simplest form of the theory. The basic law of evolution is given by

$$\left[\frac{\hbar}{\imath}\frac{\partial}{\partial t} + H\left(q, \frac{\hbar}{\imath}\frac{\partial}{\partial q_k}\right)\right]\psi(q,t) = 0, \tag{4}$$

where $\psi(q,t)$ is a complex-valued function that is an element of a Hilbert space with inner product $(\phi, \psi)$. In order to solve the Schrödinger equation (4) for the state $\psi(q,t)$, the initial state $\psi_0(q) = \psi(q,0)$ at time $t = 0$ must be specified. There are different answers to the question of which object is described by the state $\psi$. However, it is undisputed that $\psi$ is defined on the configuration space of a single particle. The predictions of QT, like the predictions of PM, are statistical in nature but much more complicated: If $a, \phi_a$ are eigenvalues and eigenfunctions of an operator $\hat{A}$, then the probability of measuring the value $a$ in the state $\psi$ is given by $|(\phi_a, \psi)|^2$.The most important difference to PM, however, is that in the evolution equation (4) of QT there are no particle trajectories that implicitly determine the probabilistic output for arbitrary times $t$, as is the case with PM. In comparison to PM, not only the initial values but also the *laws* to which the particles are subject are undetermined or non-existent. In this sense, QT is an indeterministic theory, in contrast to the cryptodeterministic theory PM.

In an earlier work by this author, three types of physical theories were introduced with regard to the role of probability [36]. Theories of type 1 are deterministic (ordinary differential equations as equations of motion without any influence of statistical uncertainty). Theories of type 2 are cryptodeterministic (initial conditions are not exactly known, but particle motion is still determined by the deterministic laws of type 1). Theories of type 3 are indeterministic (there are no longer any deterministic equations of motion for individual particles; only probabilities can be assigned). The three theories mentioned above are obviously representatives of these three classes.

Both CM and PM are classical theories. The occurrence of probabilities for initial values in PM is not regarded as problematic, as this only describes the everyday incompleteness of our knowledge. Which of the two theories is better suited as a starting point for a reconstruction of QT? There is a historical preference for the deterministic theory CM, which is deeply rooted in our thinking. However, the cryptodeterministic theory PM is more general and more realistic than CM because there are no experiments with unlimited accuracy, as pointed out by Born [8] and others. An important fact for the reconstruction of QT is the structural similarity between PM and QT: Both theories have a continuous number of degrees of freedom, so that when PM is chosen, the first of the two questions posed by Eq. (1) is already answered. For this reason, as well as for the reasons mentioned above, it is appropriate not to skip the intermediate type 2 theory on the way from the type 1 theory CM to the type 3 theory QT, and thus to choose PM as the effective starting point for the construction of QT. With this choice, the quantization method of the individuality interpretation, which links the variables of CM with those of QT according to (1) becomes meaningless. The strategy becomes a completely different one. The basic equations of QT and the form of the quantum mechanical operators are *derived* from the equations of PM. The quantization rules (1) are themselves result of this derivation.

The present paper is the third in a series of works which are based on the ensemble point of view; the previous works will be referred to as I [31] and II [32]. Of course, there are numer-ous works in the literature that take a similar view [2, 65, 7, 6, 58, 25, 64, 47, 5, 67, 52, 23, 35, 61, 36, 30, 51]. In I, II and the present theory, an attempt was made to derive QT consistently by starting with the basic equations of a physically well-understood theory and, building on this, implementing a number of clearly defined and physically meaningful postulates. The first postulate is that it must be possible to derive QT from the theory of classical statistical ensembles. From this most fundamental assumption and the fact that QT uses space-time variables as independent variables, most of the subsequent steps in the derivation of QT follow more or less automatically. The two most important steps are a linearization and a projection from phase space to configuration space. In I, II the linearization was performed first. The theory reported in II was named "Hamilton-Liouville-Lie-Kolmogorov theory" (HLLK) quoting some of the great scientists who made this work possible. We continue to use this acronym for the present theory, which is based on the same basic assumptions as I and II, but uses a different method that provides new insights.

In the present and the following work, to be referred to as IV [29], we carry out the transition from PM to QT by simply reversing the order of the above two steps. After an introduction to the general structure of the HLLK in section 2, the transition from phase space $\Omega$ to configuration space $\Sigma$ is carried out in section 3. This is done by assigning a momentum field $M(q,t)$ to each point $q \in \Sigma$ at each time $t$. There is no such field in I, II; the present approach allows a more accurate description, insofar as it provides insight into the structure of the momentum field. The discussion in section 4 shows that the components of $M(q,t)$ have to be replaced by functionally independent variables, the so-called Clebsch potentials [10] $S(q,t), Q_\alpha(q,t), P_\alpha(q,t)$. In this work we use a single Clebsch potential $S(q,t)$ to represent the three components of the momentum field of a single particle. In IV a complete representation is given which also takes into account the spin of the single-particle ensemble. In the present paper we cover only the very basics of spin theory, as our primary aim here is to understand the interface between classical physics and QT. The essential properties of this interface as well as related questions of interpretation become much more transparent when spin is neglected. In section 5, using the probability density $\rho(q,t)$ and the Clebsch potential $S(q,t)$ as dynamic variables we obtain a theory which will be referred to as "quasi-quantal approximation" (QA) of PM. This theory is closely related to the well-known quasi-classical approximation of the Schrödinger equation for spin-less particles. Section 6 discusses the unstable structure of QA, which requires consolidation. In the following two sections 7, 8 we carry out the transition from QA to QT by means of two methods, a linearization (section 7) and and a randomization (section 8).The three main steps (projection, elimination of redundancy, and linearization) are shown schematically in Figure 1. The relevant variables are shown in the rectangles, whereby the Clebsch potentials $Q_\alpha$ and $P_\alpha$ have also been taken into account here for the sake of completeness. The terms Global, Local, Not Even Local refer to the range of validity of the particle concept.

In section 9 we investigate how the concept of integral invariants changes during the transitions from PM to QT. In section 10 we return to the fundamental question of which objects are described by QT. Related questions of the transition between classical physics and QT are also discussed; among other things, we clarify the meaning of Dirac's mysterious "families of solutions". The last two sections 11 and 12 provide further discussions and concluding remarks.

## 2 Brief overview of the HLLK

Let us briefly recall the ideas on which works I, II are based. In contrast to previous ensemble theories in configuration space, see e.g. [52, 35, 30, 22], the works I, II start from the most fundamental level, namely from phase space. A classical statistical ensemble in phase space is defined as the uncountable set of all solutions of the canonical equations (2). The state of each member of the ensemble, at each instant of time $t$, is described by $2n$ numbers, namely $n$ coordinates $q = q_1, ..., q_n$ and $n$ conjugate momenta $p = p_1, ..., p_n$. The dependence of the state $q, p$ on time $t$ is ruled by the time-independent Hamiltonian $H(q,p)$. The solutions of (2) are written in the form

$$q_k = Q_k(t, t^0, q^0, p^0), \quad p_k = P_k(t, t^0, q^0, p^0), \quad (5)$$

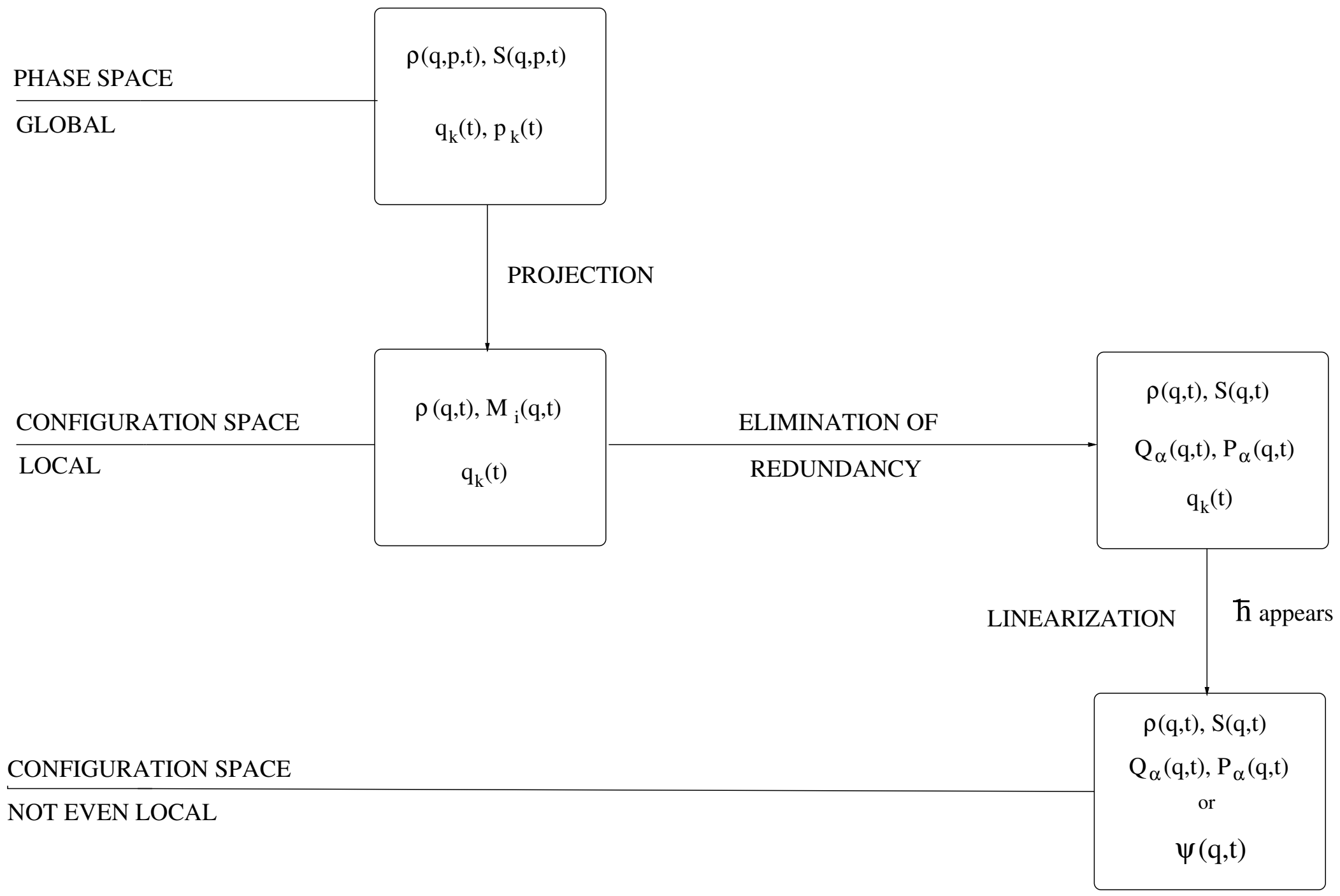

Figure 1: Schematic representation of the three main steps that lead from PM to QT. Also shown are the relevant variables and the decreasing range of validity of the particle concept.

where $(q^0, p^0)$ is the state at the initial time $t^0$; the dependence on $t_0$ will often be suppressed. The ensemble is given by the infinite set of solutions (5) with $(q^0, p^0)$ taking all possible values in the $2n$-dimensional phase space $\Omega = \mathbb{R}^n_q \times \mathbb{R}^n_p$.

The variables $q, p$ are "Lagrangian coordinates" describing properties of *particles*. Such a "particle-like" description of an ensemble is unwieldy. In the more convenient "Eulerian description", which corresponds to the usual field-theoretical formulation, the same symbols $q$, $p$ are used to denote points of a *space*, in this case points of $\Omega$. The distinction is important despite the fact that the mathematical range of values is the same in both cases. Arbitrary "Lagrangian" functions, describing the distribution of certain particle properties may be transformed to "Eulerian" functions describing corresponding fields. This is a general method [9] frequently applied in 3−dimensional fluid mechanics [3] and applied here to $2n-$dimensional phase space; in section 6 we will try to apply it to $n-$dimensional configuration space $\Sigma = \mathbb{R}^n_q$.

The mathematical realization of the Lagrangian to Eulerian transition is very simple. It is based on the assumption that the solutions $q$, $p$ of the equations of motion can be solved with respect to the initial values $q^0$, $p^0$. The following equations therefore apply

$$q = Q(t, q^0, p^0), \quad p = P(t, q^0, p^0), \tag{6}$$
$$q^0 = Q^0(t, q, p), \quad p^0 = P^0(t, q, p), \tag{7}$$

if we omit the indices for the sake of simplicity. By definition, the initial values are 'constants of the motion' if they are considered as a function of the coordinates $q(t), p(t)$ and the time $t$, i.e. they obey the differential equations

$$\begin{aligned} \frac{\partial Q^0}{\partial t} + \frac{\partial Q^0}{\partial q}\dot{q} + \frac{\partial Q^0}{\partial p}\dot{p} = 0, \\ \frac{\partial P^0}{\partial t} + \frac{\partial P^0}{\partial q}\dot{q} + \frac{\partial P^0}{\partial p}\dot{p} = 0 \end{aligned} . \tag{8}$$

Let us now consider an arbitrary function $F^0$ of the initial values $q^0, p^0$, which may also depend explicitly on $t$. If we represent the initial values with the help of Eq. (7) as a function of the coordinates $q(t), p(t)$ and the time $t$, we obtain a function

$$F(q, p, t) = F^0(Q^0(t, q, p), P^0(t, q, p), t). \tag{9}$$

Taking the fact into accout that $q, p$ do also depend on time, the (total) derivative of $F(q, p.t)$ with respect to $t$ is given by

$$\frac{\partial F}{\partial t}+\frac{\partial F}{\partial q}\dot{q}+\frac{\partial F}{\partial p}\dot{p}=\frac{\partial F^0}{\partial Q^0}\left(\frac{\partial Q^0}{\partial t}+\frac{\partial Q^0}{\partial q}\dot{q}+\frac{\partial Q^0}{\partial p}\dot{p}\right)+\frac{\partial F^0}{\partial P^0}\left(\frac{\partial P^0}{\partial t}+\frac{\partial P^0}{\partial q}\dot{q}+\frac{\partial P^0}{\partial p}\dot{p}\right)+\frac{\partial F^0}{\partial t}. \tag{10}$$

The first two terms on the right-hand side of (10) vanish as a consequence of Eq. (8). Finally, we replace the time derivatives $\dot{q}, \dot{p}$ by the velocity fields on the right-hand sides of the equations of motion (2) and obtain the general partial differential equation for the Eulerian field variable F(q,p.t):

$$\frac{\partial F}{\partial t}+\frac{\partial F}{\partial q}\frac{\partial H}{\partial p}-\frac{\partial F}{\partial p}\frac{\partial H}{\partial q}=\frac{\partial F^0}{\partial t}. \tag{11}$$

With this last step, properties of particles are described by field functions. The task now is to select suitable fields.

If we denote the probability distribution at time $t=0$ by $\rho^0(q^0,p^0)$, the Eulerian equivalent $\rho(q,p,t)$ of this time-independent Lagrangian variable, constructed according to the above scheme, fulfils Liouville's equation (3). This is the most important dynamic variable of PM, describing the distribution of possible trajectories in phase space. Liouville's equation is the counterpart of Newton's law [or its reformulation (2)] for probabilistic ensembles of trajectories. If $\rho(q,p,t)$ is known, all of the observable output (expectation values) of such a probabilistic theory may be calculated.

It is known that the solutions of the Liouville equation (3) can be constructed if the solutions of (2) are known. Nevertheless, the differences between PM and CM are important, especially for the reconstruction of QT. Observable results in a state of CM are given by values of observables $A(q,p,t)$, observable results in a state of PM are given by expectation values

$$\bar{A}_\rho(t)=\int \mathrm{d}q\,\mathrm{d}p\,\rho(q,p,t)A(q,p,t). \tag{12}$$

Let us also make a brief comment on the limit of individual particles in the ensemble theory PM. The analogue to the individual trajectories of CM are particle-like solutions of the Liouville equation, where the probability is concentrated entirely on the trajectories. The question arises as to whether the probabilistic description and the description with individual trajectories are equivalent in this limit. It turns out that this is *not* the case. We refer to the literature for a discussion of this interesting limiting case [27].

The dynamical variable $\rho(q,p,t)$ is all we need in order to calculate time-dependent expectation values of arbitrary classical observables $A(q,p,t)$; no further dynamical variables are required, in principle, to define a "state" of PM. Nevertheless, we introduce, as in I,II, the classical action on phase space as a second dynamic variable. The Lagrangian action variable $S_0^{(P)}$, defined by

$$S_0^{(P)}(q^0,p^0,t)=\int_{t_0}^{t}\mathrm{d}t' L\left(Q(t',q^0,p^0),\dot{Q}(t',q^0,p^0),t'\right), \tag{13}$$

is evaluated at the real paths $q(t')$ and depends explicitly on time $t$. The Lagrangian to Eulerian transition, constructed according to the above scheme, leads to the differential equation

$$\frac{\partial S^{(P)}}{\partial t}+\frac{\partial S^{(P)}}{\partial q_k}\frac{\partial H}{\partial p_k}-\frac{\partial S^{(P)}}{\partial p_k}\frac{\partial H}{\partial q_k}=\bar{L}, \tag{14}$$

for the Eulerian action variable $S^{(P)}(q,p,t)$. Eq. (14) will be referred to as action equation. The quantity $\bar{L}$ (the Lagrangian) is defined by

$$\bar{L}=\bar{L}(q,p,t)=p_k\frac{\partial H(q,p)}{\partial p_k}-H(q,p), \tag{15}$$

since we are only considering time-independent Hamiltonian functions here. The dynamic variable $S^{(P)}(q,p,t)$ does not play a decisive role in the formulation of the basic laws of PM, but is of essential importance for the transition to QT.

The left-hand-sides of (3) and (14) may be interpreted as total derivatives with respect to time, with a velocity field $V=(\frac{\partial H}{\partial p},-\frac{\partial H}{\partial q})$. The action variable $S^{(P)}(q,p,t)$ describes the deterministic behaviour of the entire ensemble in phase space. Note that there is no coupling between the basic differential equations (3) and (14); in particular $\rho(q,p,t)$ does not occur in the differential equation (14) for $S^{(P)}(q,p,t)$. This is a triviality in classical probabilistic physics: the dynamical laws remain intact while all the uncertainty is due to the initial states. Here we see the motivation (or one of the motivations) for introducing $S^{(P)}(q,p,t)$: Adding the law (14) to PM offers the possibility to construct an indeterministic theory (like QT) by introducing a *coupling* between $S^{(P)}(q,p,t)$ and $\rho(q,p,t)$ which destroys the concept of particle trajectories.

In I and II Schrödinger's equation was derived from PM by means of two simple steps. In the first step, the state variable $\rho$ was replaced by a complex-valued state variable $\psi$ which depends in a suitable manner (Madelung form) on $\rho$ and $S^{(P)}$. The resulting evolution equation for $\psi$ is the counterpart of Schrödinger's equation in phase space. In the second step, this equation was projected onto $\Sigma$ . This leads to the standard quantization rules and to Schrödinger's equation. Furthermore, in II the structures described by Eqs. (2), (3), (14) have be defined in an analogous way not only for the Hamiltonian function $H(q,p)$ but for any observable $A(q,p)$. In this way the most important relations of QT, namely the form of operators, Schrödinger's equation, eigenvalue equations, commutation relations, expectation values, and Born's rule have been derived in II. The simplicity of the derivation in II is a consequence of the fact that mathematical structures exist in phase space which are very similar to those of quantum theory; these structures belong to the framework of PM and are "invisible" if CM is considered as the classical counterpart of QT.

Let us extract now from this overview the general structure of HLLK. We start from PM, which is a theory of ensembles of point particles with $2n+1$ independent variables $q,p,t$, and construct a theory of type QT, with $n+1$ independent variables $q,t$. Such a construction must contain at least two essential steps, namely

- a linearization, and
- a projection from $\Omega$ to $\Sigma$.

Let us discuss these two requirements in more detail.

- The property of *linearity* for a fundamental law of physics (this means: a law containing only fixed parameters like mass and universal constants like Planck's constant, the velocity of light, or the unit of charge) is a necessary requirement if we want this law to describe a large number of well-defined processes. This follows from the fact that almost all solutions of nonlinear evolution equations become singular outside some (small) time interval. We shall use the term "locally valid" to characterize this restricted validity of solutions of nonlinear equations. On the other hand, linearity implies that all solutions of the initial value problem exist not only locally but on a global scale; this is of course a consequence of the superposition principle, which facilitates enormously the construction of solutions with prescribed initial values. We shall use the term "globally valid" to characterized the almost unrestricted validity of solutions of linear equations. The field equations of all free particles in the standard model of elementary particles are linear. When constructing field equations, the requirement for linearity is often not even mentioned.

- The requirement of *having space-time coordinates as independent variables* is also very reasonable. Space coordinates are more fundamental than momentum coordinates; the latter can only be determined if a continuous set of the former is available. All fundamental fields of physics represent properties which vary as a function of space-time and not as a function of space-momentum-time (if several particles are considered each one may have its own space coordinates). There is of course an infinite number of coordinate systems in $\Omega$ and $\Sigma$ , all of which are mathematically equivalent. However, the systems $q,p,t$ in $\Omega$ and $q,t$ in $\Sigma$ are the simplest and the only ones that allow intuitive access.

In the version of HLLK reported in I and II the linearization was performed first, followed by a projection to $\Sigma$ as a second step. As a result, in II we were able to derive the most important relations of QT very quickly. The price we have to pay for this quick derivation is that we do not gain insight into the structure of the momentum field. In this work and the following IV, we remedy this disadvantage by changing the order of the two steps.

## 3 Projection to configuration space

In this section our basic equations (2), (3), (14) will be projected to configuration space. This first step of HLLK means defining appropriate momentum fields which replace the independent variables $p$. The phase space variables $\rho(q,p,t)$ and $S^{(P)}(q,p,t)$ will be replaced by appropriate configuration space variables, which depend on $q,t$ only, and the equations of motion for these new dynamical variables will be derived. The theory obtained this way is not realized in nature but plays a very important role as an intermediate step from PM to QT.

### 3.1 Projection of particle equations

We associate with each point $q$ of configuration space (at each instant of time $t$) a particle momentum by replacing the variable $p$ by a $n-$component momentum field $M$:

$$p_k \to M_k(q,t). \tag{16}$$

Thus, $2n-$dimensional phase space is projected to a $n-$dimensional surface parameterized by the configuration space coordinates $q_k$ [46]. We may also say that we restrict our considerations from now on to a $n$-dimensional subspace $\mathbb{M}$ of phase space, which is defined, suppressing time for the moment, by $\mathbb{M} = \{(q,p) \in \Omega \,|\, p = M(q)\}$. A large part of the following considerations is devoted to a study of $\mathbb{M}$ and its various transformed versions.

Of course, the momentum field $M_k(q,t)$ should be chosen in such a way that the original structure of the canonical equations (2) is preserved as far as possible. Following Rund [56] we introduce fields $V(q,p,t)$, $v(q,t)$, $h(q,t)$, defined by

$$V_k(q,p,t) = \frac{\partial H(q,p,t)}{\partial p_k}, \tag{17}$$

$$v_k(q,t) = V_k\left(q, M(q,t), t\right), \tag{18}$$

$$h(q,t) = H\left(q, M(q,t), t\right). \tag{19}$$

Then, given that the trajectory $q_k(t)$ is a solution of the first member of (2)

$$\dot{q}_k = v_k(q,t), \tag{20}$$

the corresponding momentum $p_k(t) = M_k(q(t),t)$ must be a solution of the second member of (2). This requirement leads to the following differential equation for the momentum field $M(q,t)$:

$$\frac{\partial M_i(q,t)}{\partial t} + \left[\frac{\partial M_i(q,t)}{\partial q_l} - \frac{\partial M_l(q,t)}{\partial q_i}\right] v_l\left(q,t\right) = -\frac{\partial}{\partial q_i} h\left(q,t\right). \tag{21}$$

This nonlinear partial differential equation (recall that $v$ and $h$ depend on the unknown variables) will be referred to as canonical condition. Thus, we obtain, as a result of the projection to configuration space, new equations for the particle trajectories, namely $n$ partial differential equations (21) for $M$ and $n$ ordinary differential equations (20) for the positions $q$. These new equations replace (2)

These equations can still be used to calculate trajectories. One has to perform the following steps:

1. Assign an initial value $M_0(q)$, at $t = t_0$, to the momentum field $M(q,t)$.
2. Find a solution $M(q,t)$ of (21) for this initial value.
3. Solve the ordinary differential equations (20) to obtain the trajectory coordinates $q(t)$.
4. Use these coordinates to obtain the corresponding momenta $p(t) = M(q(t),t)$.

While trajectories still exist they are only locally valid as a consequence of the nonlinearity of the dynamical equations (21). The projected equations (21), (20) and the original canonical equations (2) are therefore not equivalent. The projection onto the configuration space leads to a loss of information regarding the description of the trajectories. The changes that the concept of trajectories undergoes through projection are further discussed in section 6.

For $n = 3$ Eq. (21) agrees with the evolution equation for the velocity field of a barotropic ideal fluid; in this context it is sometimes referred to as Lamb equation [38]. In the present work we will encounter several other formal similarities between our probabilistic theory and fluid mechanics. Despite such similarities, the different physical meaning of the basic variables should be born in mind. In particular, in fluid mechanics the density $\rho(q,t)$ is a directly measurable "material" quantity. In the present theory $\rho(q,t)$ is a rather abstract quantity, a probability density which may, for example, be measured by performing a large number of individual experiments with arbitrary large time-intervals in between.

### 3.2 Projection of ensemble equations

In this section we project the Liouville equation (3) and the action equation (14) to the subspace defined by a solution $M(q,t)$ of the canonical condition (21). The resulting partial differential equations represent an essential step on our way from PM to QT.

The projection of the Liouville equation to configuration space is a standard process which is, in a three-dimensional fluid-dynamical context, sometimes referred to as "hydrodynamical substitution". We rewrite (3) as a continuity equation in phase space:

$$\frac{\partial \rho(q,p,t)}{\partial t} + \frac{\partial}{\partial q_k}\rho(q,p,t)\frac{\partial H(q,p,t)}{\partial p_k} - \frac{\partial}{\partial p_k}\rho(q,p,t)\frac{\partial H(q,p,t)}{\partial q_k} = 0, \tag{22}$$

and restrict $\rho(q,p,t)$ according to $\rho(q,p,t) = \rho(q,t)\delta(p-M(q,t))$, where $\rho(q,t)$ is the new probability density in configuration-space, and $\delta$ is the $n$-dimensional delta function. Performing the integration with respect to the momentum variables leads to the desired continuity equation in configuration-space

$$\frac{\partial \rho(q,t)}{\partial t} + \frac{\partial}{\partial q_k}\rho(q,t)V_k(q,M(q,t),t) = 0, \tag{23}$$

which may of course be interpreted as a probabilistic conservation law with $\rho v(q,t)$ as probability current.

It is also useful to consider the projection of the action equation (14) to configuration space. We define a configuration space field $s(q,t)$ according to

$$s(q,t) = S^{(P)}(q,M(q,t),t). \tag{24}$$

We omit an index $M$ but bear in mind that $s(q,t)$ depends by definition on the momentum field $M$. We may replace in (14) the partial derivatives of $S^{(P)}(q,p,t)$ with respect to $q$ and $t$ by partial derivatives of $s(q,t)$ with respect to the same variables. After performing several such manipulations, Eq. (14) takes the form

$$\frac{\partial s}{\partial t} - \left.\frac{\partial S^{(P)}}{\partial p_k}\right|_{p=M}\left[\frac{\partial M_k}{\partial t} + \left(\frac{\partial M_k}{\partial q_i} - \frac{\partial M_i}{\partial q_k}\right)v_i + \frac{\partial}{\partial q_k}h\right] + v_k\left(\frac{\partial s}{\partial q_k} - M_k\right) + h = 0. \tag{25}$$

As $M_k(q,t)$ is a solution of (21) the bracket vanishes and Eq. (25) takes the form

$$\frac{\partial s}{\partial t} + v_k\left[\frac{\partial s}{\partial q_k} - M_k(q,t)\right] + h = 0. \tag{26}$$

This is the projection of Eq. (14) to configuration space derived in a straightforward way. Starting from the phase space action (13), the derivation of (26) was performed in two steps. The first step was a Lagrangian to Eulerian transition in phase space leading to Eq. (14). The second step was a projection to configuration space. If these two steps are carried out in reverse order, one obtains of course the same result (26). Clearly, this equation takes the form of the Hamilton-Jacobi equation when $M$ is given by the gradient of $s$.

Starting from the canonical equations (2), we performed two processes which generated a number of variables, which we will now list again for the sake of clarity. With the help of the Lagrangian to Eulerian transition we have generated the field variables $\rho(q,p,t)$ and $S^{(P)}(q,p,t)$ whose behaviour is still determined by the particle variables $q_k(t)$, $p_k(t)$. On the other hand, we have projected the canonical equations directly onto the configuration space, generating momentum fields $M_k(q,t)$ that obey the canonical condition (21), as well as unstable particle variables, which we continue to denote by $q_k(t)$ for the sake of simplicity. The projection of the field variables $\rho(q,p,t)$ and $S^{(P)}(q,p,t)$ with the help of the $M_k(q,t)$ leads to the fields $\rho(q,t)$ and $s(q,t)$. As a result, we have $n+2$ fields $M_1(q,t),...M_n(q,t),\rho(q,t),s(q,t)$ in the configuration space, in addition to the unstable particle variables. The associated field equations form a *classical* theory, which means that the probability density $\rho(q,t)$ does not appear in the equations ruled by the motion of the particles. These "pure" particle (field) equations are the canonical condition (21) and the generalisation of the Hamilton-Jacobi equation (26). Since the phase space variable $S^{(P)}(q,p,t)$ is a redundant quantity (with regard to the description of particle motion), it is reasonable to assume that this also applies to its projection $s(q,t)$. We will show, considering irrotational momentum fields in section 5, that there is indeed a relation between $M_1(q,t),...M_n(q,t)$ and $s(q,t)$. This means that in the following we can interpret equations (23) and (21) for the variables $\rho(q,t)$ and $M_k(q,t)$ as our fundamental system of field equations.

The projection of the canonical equations (2) to configuration space has led us to the particle Eqs. (20) and the field Eqs. (23) and (21). These relations define a theory which suffers from two shortcomings. The first, mentioned already in section 3.1, is given by the fact that trajectories are only locally well-defined. The second is given by the fact that the $n$ components $M_1(q,t),..,M_n(q,t)$ of the momentum field may not be *functionally independent* from each other.

## 4 Introducing potentials

In this section we show that the next step in the development of the HLLK has to be a replacement of the dynamical variables $M_i$ by potentials. First we consider arbitrary particle numbers $N$, but later specialize to $N=1$ and to an irrotational momentum field, which may be described by a single scalar potential $S$. The general (correct) treatment of the case $N=1$ requires 3 independent potentials and will be reported in IV.

If we consider a system of $N$ free particles in phase space, the $3N$ components of the particle momentum are not subject to any restrictions. It seems unlikely that this freedom will be maintained for the $3N$ fields $M_k$ after projection to $\mathbb{M}$. To examine this question in a more definite way, we need a mathematical theory that tells us how to eliminate redundancy from a set of fields. This question was studied by the mathematician Pfaff in 1815 (see [24] for references and background information). Pfaff's problem may be formulated as follows:

> Given $n$ fields $M_1(q,t),...,M_n(q,t)$ not necessarily functionally independent from each other. What is the minimal number $L$ of fields $F_1(q,t),..,F_L(q,t)$ required to derive the $M_i(q,t)$, and which relations exist between the fields $M_i(q,t)$ and the "potentials" $F_i(q,t)$ ?

This problem is equivalent to a "complete reduction" of the $1-$form, or Pfaffian form $\omega = M_i(q)dq_i$ i.e. to a transformation of the Pfaffian form into an equivalent form with a minimal number of terms. The number $L$ is referred to as "class" of the vector field $M=\{M_1,...,M_n\}$. A detailed treatment of Pfaff's problem may be found, e.g., in Caratheodory's book [9].

The number $L$ may be determined (by finding he rank of a certain matrix [9]) if the functional form of $M(q,t)$ is known. As $M(q,t)$ is not known to us we cannot take advantage of this mathematical result. Rather, we have to treat $L$, or the integer $m$ introduced below, as a kind of "adjustable parameter" (taking of course logical and physical consistency requirements into account). The part of Pfaff's solution which is most interesting for us is the form of the expansion. For momentum fields of odd class there is an integer $m$ given by $L=2m+1$. In this case $M$ may be written in the form

$$M_k(q,t) = \frac{\partial S(q,t)}{\partial q_k} + P_\alpha(q,t)\frac{\partial Q_\alpha(q,t)}{\partial q_k}, \qquad (27)$$

where $S(q,t)$, $P_\alpha(q,t)$, $Q_\alpha(q,t)$ are $2m+1$ *independent* functions of $q_1,..,q_n,t$ (Greek indices $\alpha$, $\beta$,.. run from 1 to $m$ and double occurrence of these indices entails a summation from 1 to $m$). The functions $S(q,t)$, $P_\alpha(q,t)$, $Q_\alpha(q,t)$ are referred to as Clebsch potentials [10]. For fields of even class there is an integer $m$ given by $L=2m$ and $M$ may be written in the form

$$M_k(q,t) = P_\alpha(q,t)\frac{\partial Q_\alpha(q,t)}{\partial q_k}, \qquad (28)$$

with $m$ pairs of functions $P_\alpha(q,t)$, $Q_\alpha(q,t)$. We should note that the above expansions are only locally valid. Decompositions in terms of Clebsch potentials are mainly found in fluid mechanics, while in electrodynamics the Helmholtz decomposition is used almost exclusively. The great advantage of the Clebsch decomposition is the simple form of the rotational part and the simple extension to arbitrary dimensions. Its biggest disadvantage is that it is only locally valid. This is a serious limitation for many tasks in physics but not for the present one. The reason is that the global validity of the present theory has already been lost (by projecting to configuration space) *before* the elimination of redundancy is implemented.

Two questions now arise, namely (i) which one of the above two expansions is suitable for our purposes, and (ii) is there really redundancy or may we continue to use $M_k$ as our dynamical variables? Both of these questions can be answered in a simple way, just by considering the form of the expansions (27) and (28). We use the fact that the concept of constraints makes only sense in macroscopic physics. This means we can identify the coordinates of the configuration space with the Cartesian coordinates of one or several particles. In other words, if we have $N$ particles then we have $n = 3N$ spatial degrees of freedom, and a change in $n$ is always a consequence of a change in $N$ ; in particular $n = 3, 6, 9, 12, ...$ if $N = 1, 2, 3, 4, ...$

From the fact that the structure of our (future) theory can not depend on the number of particles, we conclude first of all that only one of the two representations (27) and (28) can be true for all $N$. Then, only (27) survives. In fact, the representation (28) leads to an even number of fields and fails to describe the single particle case, $N = 1$, which requires three fields. Thus, in the three-dimensional space of nonrelativistic physics, which is considered throughout the present study, we have to use the odd representation, Eq. (27), with non-vanishing Clebsch potential $S(q, t)$ . On the other hand, for a covariant formulation in the four-dimensional space-time of relativistic physics the even representation (28) with $S(q, t) = 0$ is appropriate, as has been shown by Rund [56], [57].

To answer the second question we first assume that there is no redundancy, that is, $L = n = 3N$. Since $L$ is alternately odd and even with increasing N, this assumption leads to a contradiction. Thus, there must be redundancy, at least for some $N$ the number $L$ of independent components of $M$ must be less than $n$. This leads to the important conclusion that we have to use *potentials* as dynamic variables instead of momentum fields. We have derived a fundamental structural property of QT, whose origin is seldom asked for.

We know now that only (27) can be used, i.e. the class is given by $L = 2m + 1$. Secondly, we know that $L$ cannot be equal to $n$ for all $n$; i.e. a relation of the form $L = n - k$, where $1 \leq k \leq n - 1$, will exist at least for some $n$ ($k$ is the number of functional relations between the $M_i$) . Combining both relations we have

$$n - k = 2m + 1. \tag{29}$$

Since $n, k, m$ are all integers we have a countable number of solutions for $k$ and $m$. There is a solution of maximal redundancy for arbitrary $n$ which is given by $m = 0$ and $k = n-1$. For $N = 1$ there is no redundancy, the relation $3-k = 2m+1$ can only be true for $k = 0$ and $m = 1$ (excluding here the case of maximal redundancy). For $N = 2$ we have necessarily redundancy; the relation $6-k = 2m+1$ can only be true for nonzero $k$; the possible pairs $(k, m)$ are from the set $\{(1, 2), (3, 1)\}$. For $N = 3$ the possible pairs $(k, m)$ are from the set $\{(0, 4), (2, 3), (4, 2), (6, 1)\}$, etc. Inspecting the various possible solutions for each $N$ one finds that only a single "regular" solution for $k$ and $m$ exists, namely the linear solution $k = N - 1$, $m = N$. Only for this "regular" solution the increase in the total number $2m + 2$ of dynamical variables (taking $\rho$ into account) is the same for all $N$, namely 2. For all other solutions this number changes in an irregular way; examples for such solutions (for $N = 1, 2, 3, 4, ..$) are $2m + 2 = 4, 4, 6, 4, ..$ or $2m + 2 = 4, 6, 4, 8, ...$ These irregular solutions, are, although not impossible from a logical point of view, obviously unacceptable, and can be excluded. Let us summarize the results of our analysis for $n = 3N$.

- To expand the momentum fields $M_1, .., M_n$ only the odd representation, Eq. (27) can be used, i.e. $L = 2m+1$. As regards the number $m$ two possibilities exist:

- Either the case of maximal redundancy is realized, i.e. $m = 0$ for all $N$. In this case the momentum field $M$ is the gradient of a scalar function $S(q_1, .., q_n)$.

- Or $m = N$ and $N-1$ relations exist among the $3N$ momentum field components. In this case $M$ can be expanded in $2m+1$ Clebsch potentials, where $m = N$. Thus, a $N-$particle ensemble is described by a function $S$ and $N$ pairs of dynamic variables $P_\alpha$, $Q_\alpha$.

The most important conclusion of this section is that the momentum fields *have to be replaced by potentials*. This conclusion is only correct for theories with a variable number of particles. For theories like fluid mechanics, where the number of degrees of freedom is always 3 (corresponding to $N = 1$) both the three components of the momentum (or the velocity) and the three Clebsch potentials may equivalently be used as dynamic variables. For $N = 1$ it is also possible to rewrite QT with the help of both types of variables [63, 26]. However, for our present task of reconstructing QT for arbitrary $N$, only potentials are appropriate.

## 5 Basic equations for irrotational momentum fields

In the present work we restrict ourselves to the case $m = 0$, where only a single potential $S$ occurs. The case $m = 1$, which explains the spin of a single particle, will be treated in IV. It is useful to introduce a quantity which characterizes the purely rotational, or *vortical* part of the momentum field $M$. An appropriate quantity is the vorticity tensor $\Omega_{ik}$, defined by

$$\Omega_{ik}(q,t) := \frac{\partial M_k(q,t)}{\partial q_i} - \frac{\partial M_i(q,t)}{\partial q_k}. \qquad (30)$$

One can show that the tensor $\Omega_{ik}$ *moves with the flow* defined by the particle equations of motion; in particular, $\Omega_{lj}(q,0) = 0$ implies $\Omega_{ik}(q,t) = 0$ for all future times [56].

A momentum field obeying $\Omega_{ik} = 0$ is referred to as an irrotational momentum field; this happens if either the $P_\alpha(q,t)$ or the $Q_\alpha(q,t)$ are constant. This case corresponds to $m = 0$ and $M$ is, according to (27), given by the $n$-dimensional gradient of a single Clebsch potential $S$,

$$M_k(q,t) = \frac{\partial S(q,t)}{\partial q_k}. \qquad (31)$$

For such a field $n-1$ functional relations exist between the $M_i$, no matter how large the number of particles. A physical reason for such strong dependencies between the momentum components (such a strong redundancy) cannot be seen. In particular, it would be reasonable to assume that for $N = 1$ the number of independent components is equal to the maximal possible number 3, as there is *no restriction* in principle for momenta in three-dimensional space. In spite of this unsatisfactory situation, which will be further discussed and resolved in IV, the case of the irrational momentum field is very important because it leads to Schrödinger's equation.

For irrotational momentum fields the change of dynamical variables from $M_i$ to $S$ in the field equations is very simple. As a consequence of Eq. (31) the bracket in the canonical condition (21) vanishes. The latter reduces to the Hamilton-Jacobi equation if an unspecified function $f(t)$ is eliminated by means of a redefinition of $S(q,t)$. If we add the particle equation (20) and the continuity equation (23), we obtain the three basic differential equations

$$\dot{q}_k = V_k(q, \frac{\partial S(q,t)}{\partial q}, t), \qquad (32)$$

$$\frac{\partial S}{\partial t} + H(q, \frac{\partial S}{\partial q}, t) = 0, \qquad (33)$$

$$\frac{\partial \rho(q,t)}{\partial t} + \frac{\partial}{\partial q_k}\rho(q,t) V_k(q, \frac{\partial S}{\partial q}, t) = 0. \qquad (34)$$

Thus, the irrotational case may be described by only two dynamical field variables, $S(q,t)$ and $\rho(q,t)$ no matter how large $n$. These equations have to be solved as initial value problems; the steps required to calculate trajectories are similar to the ones listed in section 3.1.

In section 3.2 we noted that there must be a functional relationship between the projected action variable $s(q,t)$ and the momentum fields $M_1(q,t), .., M_n(q,t)$. In the special case $m = 1$ considered at the moment, this means that there must be a relationship between $s(q,t)$ and the Clebsch potential $S(q,t)$. This relationship, which also provides information about the physical meaning of the variable $S(q,t)$, may easily be found by comparing (33) with the projected action equation in the form (26). According to the definition (24) the projected action variable $s(q,t)$ is now given by

$$s(q,t) = S^{(P)}\left(q, \frac{\partial S}{\partial q}, t\right). \qquad (35)$$

Inserting (31) in the projected action equation (26) and using (33) shows that $S(q,t)$ differs from $s(q,t)$ by a function which is constant along the trajectories defined by the velocity field $v(q,t) = V(q, \frac{\partial S}{\partial q}, t)$:

$$\frac{\mathrm{D}}{\mathrm{D}t}(s - S) = 0, \text{ where } \frac{\mathrm{D}}{\mathrm{D}t} = \frac{\partial}{\partial t} + v_k \frac{\partial}{\partial q_k}. \quad (36)$$

The symbol defined by the second member of Eq. (36) is referred to as total derivative. Thus, the Clebsch potential $S(q,t)$ is closely related to the projected action variable $s(q,t)$; the simplest solution of (36) is the trivial solution $s - S = 0$.

Equations (33) and (34) represent the basic equations of the "quasi-quantal approximation" (QA) to PM. We use this term because (33) and (34) have been obtained here, in the framework of HLLK, as a first step towards QT starting from the classical theory PM. A linearization of these equations leads to Schrödinger's equation, as will be shown in section 7. The same equations (33) and (34) constitute also the basis of the so-called quasi-classical approximation of QT. Although both theories share the same basic equations, they differ in the additional conditions that also decisively determine the form of the solutions. In the QA the differential equations are to be solved with given initial values, as is the case in QT. In the quasi-classical approximation the solutions are mostly constructed with the help of complete solutions of the Hamilton-Jacobi equation using a phase space method that goes back to van Vleck [65]; further comments on the complete solutions method will be given in section 10.3.

## 6 The structure of the quasi-quantal approximation

The basic equations (32), (33), (34) of the QA represent the result of the first of the two essential steps of the HLLK. They describe the simplest version of an ensemble theory projected onto configuration space $\Sigma$. Let us ask now whether or not this theory makes sense from a physical point of view. If not, we ask what further steps are required to construct a physically meaningful theory in $\Sigma$.

Our starting point was the physically meaningful theory PM of statistical ensembles in phase space $\Omega$. The fact that this theory makes physical sense is reflected in its mathematical structure. The Cauchy problem for systems of first-order ordinary differential equations is well-defined for rather weak mathematical conditions [1]. This implies that (almost) all solutions of PM are globally valid and that, consequently, these solutions define an invertible map

$$l_t : \Omega \to \Omega, \quad q = Q(t, q_0, p_0), \; p = P(t, q_0, p_0), \quad (37)$$

of phase space onto itself. Actually, a continuous family of maps (37), parametrized by the time $t$, exists corresponding to the fact that, at each instant of time, each point of phase space belongs to a single trajectory. The existence of this family implies the validity of the Liouville equation (3) which is, like all fundamental laws of nature, a *linear* differential equation. Linearity is a necessary and sufficient condition for "global validity" of a physical law as mentioned already in section 1. The global validity of the Liouville equation and the global validity of the canonical equations are closely related to each other.

During the transition from PM to QA the manifold of independend variables has been reduced from $\Omega$ to $\Sigma$. An invertible map, analogous to (37) but now for the reduced manifold of QA, would be given by

$$l_t^c : \Sigma \to \Sigma, \quad q = Q^c(q_0, t). \quad (38)$$

If such a map existed, it would imply a globally valid, linear "Liouville equation" in configuration space, which would then be the universal law, we were looking for. Such a map does not exist, as mentioned already. From the solution of the initial-value problem of the Hamilton-Jacobi equation (33) the associated momentum field $M(q,t)$, with initial value $M_0(q)$, may be calculated with the help of relation (31). Then, particle trajectories $q(t)$, $p(t) = M(q(t),t)$ on the (irrotational) momentum surface $\mathbb{M}$ are obtained as solutions of Eqs. (32). These trajectories are locally well-defined, in a certain vicinity of $t_0$, but the momentum surface $\mathbb{M}$ defined by $M_0(q)$ immediately starts deforming during its movement through phase space. As a consequence, the map from configuration space to the (deformed) momentum hypersurface $\mathbb{M}$ becomes multivalued and the concept of trajectories breaks down in QA. This breakdown is closely related to the nonlinearity of Eq. (33). Clearly, we cannot cover the whole of phase-space by means of a single momentum surface (we would need an infinite number as in the "complete solutions" of the theory of canonical transformations).

The breakdown of the concept of particle trajectories is not related to the introduction of potentials; this also happens in the earlier theory, with hydrodynamic variables, reported in section 3. However, in the QA, as derived from QT, the singularities that occur when the concept of trajectories collapses have been carefully studied, see e.g. [55, 4, 12]. Thus, the concept of trajectories - which represents the "deterministic part" of our probabilistic theory - is still present in the QA but has been seriously "weakened", due to its loss of global validity. This is the price we have to pay for the enormous reduction in the number orf degrees of freedom brought about by the projection.

This discussion leads us to suspect that we cannot construct a globally valid probabilistic theory for particles in configuration space, while keeping at the same time the concept of particle trajectories (globally) intact. A probabilistic particle theory makes predictions of a probabilistic nature (predictions to be verified by means of statistical ensembles) about particles. This definition, which is based on the physical principle of verification by observation, does not require that deterministic laws for particle trajectories are part of a probabilistic theory. We have already discussed theories of this kind ("theories of type 3") in section 1.

This discussion also leads us to suspect that QT is a globally valid modification of the QA. In order to test if this working hypothesis is true we have to *linearize* the partial differential equations (33), (34). This will be done in section 7; an alternative way of deriving the resulting linear equations will be described in section 8. There seems to be no other way to restore the former global validity (of PM in phase space) in configuration space.

## 7 Transition to quantum theory by linearization

In this section the second essential step of the HLLK is carried out, which leads from the QA to the final QT. This step can be carried out in two very different ways. These two derivations illuminate the nature of QT from two different perspectives. In this section the transition is carried out by means of a linearization.

The linearization of the spinless version of the QA, which we consider in this work, can be carried out in a relatively simple way. As a first step we introduce a quasiquantal wave function which is defined in an analogous way as in II:

$$\psi = \sqrt{\rho} e^{\frac{\imath}{\hbar} S}. \tag{39}$$

However, the quantities and $S$ and $\rho$ satisfy now (33) and (34). The "linearizing transformation" (39) was first introduced by Madelung [44]. In order to be able to carry out a comparison with the original equations (14) and (3) in phase space, we use the explicit functional form of the Hamilton function $H(q,p) = p_k p_k/2m + V(q)$ for a single particle and write equations (33) and (34) in the following form:

$$D_t S = \bar{L}, \tag{40}$$

$$\hat{\mathbb{L}}_{\frac{1}{2}} \rho^{\frac{1}{2}} = 0, \tag{41}$$

where the Lagrangian $\bar{L}$ and the operator $\hat{\mathbb{L}}_{\frac{1}{2}}$ are defined by $\bar{L} = \frac{1}{2} v_k M_k - V(q)$ and $\hat{\mathbb{L}}_{\frac{1}{2}} = \partial_t + v_k \partial_k + \frac{1}{2}(\partial_k v_k)$. We introduced here the abbreviations $\partial_t$, $\partial_k$ for the derivatives with respect to $t$, $q_k$ and the abbreviation $D_t$ for the total derivative defined in Eq. (36). If we compare Eqs. (40), (41) with the analogous Eqs. (14), (3) in phase space, we find that the following properties that are valid in phase space no longer apply in configuration space:

- The divergence of the velocity field vanishes.
- The probability density moves with the flow.
- The operator $v_k \partial_k$ multiplied by the imaginary unit is self-adjoint.

The absence of these properties is related to the instability of the trajectories in configuration space discussed in Section 6.

In the next step we evaluate the expression $-\frac{\hbar}{\imath}\hat{\mathbb{L}}_{\frac{1}{2}}\psi$. Using the basic Eqs. (40), (41) we obtain the relation $(-\frac{\hbar}{\imath}\hat{\mathbb{L}}_{\frac{1}{2}} + \bar{L})\psi = 0$ which is equivalent to (33) and (34), and has some similarity to the generalized Koopmann-von Neumann equation derived in I. We have to transform this equation, which takes the form

$$\left[-\frac{\hbar}{\imath}\partial_t - \frac{\hbar}{\imath}v_k\partial_k - \frac{1}{2}\frac{\hbar}{\imath}(\partial_k v_k) + \frac{1}{2}v_k(\partial_k S) - V(q)\right]\psi = 0, \tag{42}$$

when written out, into an equation (not necessarily linear) for the complex variable $\psi$. According to the definition of $\psi$ it is possible to replace the derivatives of $S$ by the derivatives of $\psi$ and $\rho$ . This may be done with the help of the relation

$$(\partial_k S)\psi = \frac{\hbar}{\imath}\left[\partial_k\psi - \frac{1}{2\rho}(\partial_k\rho)\psi\right], \tag{43}$$

whose validity is easily verified with the help of the definition (39). We replace the fourth term of (42) with the help of (43) and write the third term of (42) in the form

$$-\frac{1}{2}\frac{\hbar}{\imath}(\partial_k v_k)\psi = -\frac{1}{2}\frac{\hbar}{\imath}\partial_k v_k\psi + \frac{1}{2}\frac{\hbar}{\imath}v_k\partial_k\psi. \tag{44}$$

Using (43) again, the equation of motion (42) takes the form

$$\left[-\frac{\hbar}{\imath}\partial_t + \frac{\hbar^2}{2m}\left(\partial_k + \frac{1}{2\rho}(\partial_k\rho)\right)\left(\partial_k - \frac{1}{2\rho}(\partial_k\rho)\right) - V(q)\right]\psi = 0. \tag{45}$$

A few further elementary manipulations finally lead to the nonlinear differential equation

$$\left[\frac{\hbar}{\imath}\partial_t + \hat{H}\right]\psi - \frac{\hbar^2}{2m}\frac{1}{\sqrt{\rho}}(\partial_k\partial_k\sqrt{\rho})\psi = 0, \tag{46}$$

where $\hat{H} = \frac{1}{2m}\hat{p}_k\hat{p}_k + V(q)$, $\hat{p}_k = \frac{\hbar}{\imath}\partial_k$, and $\rho = \psi^*\psi$. The "quantum-like" classical differential equation (46) for the complex variable $\psi$ is essentially equivalent to the relations (33) and (34), apart from certain uniqueness conditions [66] for $\psi$ which are not important in the present context (see however section 9 ). Let us note that (46) has been reported several times before in the literature (see e.g. [58, 54, 36, 59]), what is new here is the derivation from the probabilistic description of classical motion in phase space.

The linearization of Eq. (46) is now very simply carried out by deleting the last, non-linear term, resulting in our final result, the Schrödinger equation (4).

## 7.1 The physical meaning of the linearization

The transition from PM to QT can be represented schematically in the form of two steps: PM ⇒ QA ⇒ QT. It is the second step QA ⇒ QT, the linearization of QA, that leads to the most radical changes relative to classical physics and to the greatest difficulties in understanding QT. Let us take a closer look at what happens during the linearization.

An interesting question in this context is how the basic equations of the QA are modified in the transition to QT. A brief calculation, first performed by Madelung [44], shows that the continuity equation (34) remains unchanged while the Hamilton-Jacobi equation (33) takes the following form:

$$\frac{\partial S}{\partial t} + H(q, \frac{\partial S}{\partial q}) = \frac{\hbar^2}{2m}\frac{1}{\sqrt{\rho}}(\partial_k\partial_k\sqrt{\rho}). \tag{47}$$

Thus, the transition from QA to QT leads to an additional term $T_Q = (\hbar^2/(2m\sqrt{\rho}))(\partial_k\partial_k\sqrt{\rho})$ on the right-hand side of (33). Clearly, the term $T_Q$ introduces a coupling between the action variable $S$ and the probability density $\rho$. What does that mean exactly?

Let us recall that in the QA the action variable $S$ defines a particle momentum at every point of configuration space. The solutions $S$ of the Hamilton-Jacobi equation define, in combination with the particle equation (32), particle trajectories - even if these trajectories are only locally well-defined as discussed in section 6. Clearly, the coupling term $T_Q$ destroys this quasiclassical concept of a local particle momentum. Let us recall at this point that equations (34) and (47) are to be solved as an initial value problem. From the very beginning, all particle trajectories and momenta are "mixed together", depending on the specification of $S$ and $\rho$ at time $t = 0$. As a result, the particle equation (32) looses its meaning in QT. Section 9 confirms this conclusion in a different context. Only the two partial differential equations (47) and (34) survive the transition to QT. These two relations, which are equivalent to Schrödinger's equation (4), determine the time-development of an object, which may be described as a new kind of *ensemble where no separation of deterministic and probabilistic properties is possible.*

For this new type of ensemble, the classical distinction between states and observables no longer exists. During the transition QA $\Rightarrow$ QT, the observables of classical physics are converted into operators. These operators are often referred to as "observables". This term is a misleading by-product of the individuality interpretation. The observables of any physical theory are of course real numbers; in the case of QT the expectation values are the observable quantities. In summary, it can be said that during the transition from PM to QT a *randomization* took place - which means that the remaining elements of determinism (the laws of particle motion still present in PM, and even in QA) were eliminated.

Using the nomenclature of reference [36], we may say that, starting from PM and arriving at QT, we performed a transition from a type 2 theory to a type 3 theory. In an intermediate step, the QA emerged, which in its properties - with its reduced validity of the particle concept - can actually be classified as a theory *between* type 2 and type 3. Compared to the version of HLLK considered in I, II, the occurrence of QA allows a more precise investigation of the changes that occur during the transition from PM to QT. For example, in section 8 we will perform the randomization that takes place during the transition QA $\Rightarrow$ QT using appropriate statistical concepts.

## 7.2 Comparison of quantization rules

It is interesting to compare the role of the quantization rules in I, II with the corresponding steps in the present version of HLLK. Two steps have always to be carried out, a linearization and a projection onto configuration space. If the vertical component of the momentum field is neglected, as is done here, the order of these two steps does not matter.

Let us begin with the quantum-like classical (phase space) equation

$$\left[\frac{\hbar}{\imath}\frac{\partial}{\partial t} - \frac{\hbar}{\imath}\frac{\partial H}{\partial q_k}\frac{\partial}{\partial p_k} + \frac{\partial H}{\partial p_k}\left(\frac{\hbar}{\imath}\frac{\partial}{\partial q_k} - p_k\right) + H\right]\psi^{(c)} = 0, \tag{48}$$

derived in I, II. The classical wave function $\psi^{(c)}$ is defined as in (39) but with $\rho$ and $S$ replaced by quantities depending on $q$, $p$, $t$. The classical equation (48) is the result of a suitable linearization in phase space, which was carried out in I, II as a first step. The second step in I, II, the projection onto the configuration space, was realized by implementing the quantization rules

$$\frac{\partial}{\partial p_k} \Rightarrow 0, \quad p_k \Rightarrow \frac{\hbar}{\imath}\frac{\partial}{\partial q_k} \tag{49}$$

which may easily be read off from (48).

On the other hand, in the present version of HLLK, the quantum-like quasi-classical (configuration space) equation (46) is obtained by performing a suitable projection onto configuration space in a first step. Roughly speaking, this first step corresponds to the first of the above quantization rules (49). The second step, the linearization, only required the elimination of the nonlinear term in in (46); the second quantization rule in (49) was not explicitly used here. However, the linearization can also be performed using this second quantization condition (or its analog). We recall that in the course of the projection the variable $p$ was replaced by the field $M(q,t)$, which is given by the gradient of the field $S(q,t)$. Therefore, the second, more familiar quantization rule in (49) corresponds in the present version of the HLLK to the rule

$$\partial_k S \Rightarrow \frac{\hbar}{\imath}\partial_k. \tag{50}$$

If we apply this quantization rule to equation (43), we find that the linearization must be carried out by eliminating the gradient of $\rho$ (second term on the right-hand side of (43)). The appropriate relation to carry out the linearization in this direct way is equation (45), which again yields Schrödinger's equation (4).

## 8 Transition by randomization

The above derivation of Schrödinger's equation by linearization of QA is based on the idea that a fundamental law of nature must permit a large variety of solutions and also on the empirical fact that all field equations for free particles are linear. The resulting theory is purely probabilistic - without deterministic laws for particle trajectories. In this sense, the physical effect of the formal process of linearization may be interpreted as a randomization. In this section we try to understand this randomization on a deeper level by replacing the process of linearisation with certain statistical assumptions that lead to the same final result as the linearisation. The following developments represent an improved version of earlier work [36, 30] of the present author.

Our task is to construct a theory in which the deterministic equation (32) is no longer included, but which is otherwise as similar as possible to the basic equations (33) and (34) of QA. Therefore, we continue to use $\rho$ and $S$ as our basic variables. Furthermore, we assume that the fundamental conservation law of probability (34) remains unchanged and that the momentum field takes the same form as in the QA. Another reasonable assumption is that the particle equations of motion (2) "survive" in a statistical sense, i.e. hold true not for particle variables but for the corresponding expectation values,

$$\frac{\mathrm{d}}{\mathrm{d}t}\overline{q_k} = \frac{\overline{p_k}}{m}, \qquad \frac{\mathrm{d}}{\mathrm{d}t}\overline{p_k} = \overline{F_k}. \tag{51}$$

Here, the expectation values in configuration space $\overline{q_k}$, $\overline{p_k}$, $\overline{F_k}$ are defined as follows:

$$\overline{q_k} = \int \mathrm{d}^3q\,\rho(q,\,t)\,q_k, \tag{52}$$

$$\overline{p_k} = \int \mathrm{d}^3q\,\rho(q,\,t)\,\frac{\partial S}{\partial q_k}, \tag{53}$$

$$\overline{F_k} = -\int \mathrm{d}^3q\,\rho(q,\,t)\,\frac{\partial V}{\partial q_k}. \tag{54}$$

These are the basic and relatively weak constraints to which we subject our randomized theory.

The first of the Ehrenfest-like relations (51) results from the assumptions made and does not need to be considered further. Using the second of the relations (51) and the continuity equation (34) we derive the following integrodifferential equation:

$$\int \mathrm{d}q \frac{\partial \rho}{\partial q_k}\left[\frac{\partial S}{\partial t} + \frac{1}{2m}\sum_{j=1}^{3}\left(\frac{\partial S}{\partial q_j}\right)^2 + V\right] = 0. \tag{55}$$

In the framework of the QA, this equation is fulfilled due to Hamilton-Jacobi equation (33). We now assume that the modified Hamilton-Jacobi equation of the randomized theory differs from the Hamilton-Jacobi theory of QA only by an additional term $L_0$. We write the modified Hamilton-Jacobi equation in the form

$$\bar{L}(q,t) - L_0 = 0, \tag{56}$$

where $\bar{L}$ is defined by

$$\bar{L} := \frac{\partial S}{\partial t} + \frac{1}{2m}\sum_j\left(\frac{\partial S}{\partial q_j}\right)^2 + V. \tag{57}$$

In order for the Ehrenfest-like relations (51) to remain valid in the randomized theory, the unknown function $L_0$ must obviously satify the constraint

$$\int \mathrm{d}q \frac{\partial \rho}{\partial q_k} L_0 = 0. \tag{58}$$

We see that the statistical condition (51) is not sufficient to determine the the basic equations of the randomized theory. Further constraints of a statistical nature are required to fix $L_0$.

A statistical constraint that is in agreement with our general strategy is the following. We require that the space-time average of the difference between $\bar{L}$ and $L_0$ is minimal. This leads to the variational problem

$$\delta \int \mathrm{d}t \int \mathrm{d}q \rho \left(\bar{L} - L_0\right) = 0. \tag{59}$$

In order to write down the Euler-Lagrange equations for $\mathcal{L} = \rho\left(\bar{L} - L_0\right)$ in a compact way, we introduce the two-component field $\chi_\alpha$, with $\chi_1 = \rho$ and $\chi_2 = S$. If derivatives up to first order with respect to $t$ and second order with respect to $q_k$ are taken into account, the Euler-Lagrange equations of the variational problem (59) are given by

$$\frac{\partial \mathcal{L}}{\partial \chi_\alpha} - \frac{\partial}{\partial t}\frac{\partial \mathcal{L}}{\partial(\partial_t \chi_\alpha)} - \frac{\partial}{\partial q_k}\frac{\partial \mathcal{L}}{\partial(\partial_k \chi_\alpha)} + \frac{\partial}{\partial q_k}\frac{\partial}{\partial q_l}\frac{\partial \mathcal{L}}{\partial(\partial_k \partial_l \chi_\alpha)} = 0. \tag{60}$$

For the "correct" $L_0$, the Euler-Lagrange equations, with regard to $\rho$ and $S$, must agree with the new field equations and the function $\rho(\bar{L}-L_0)$ must agree with the Lagrangian density . On the other hand, we have already determined the form of these field equations; they are given by (56) and (34) and do *also* depend on $L_0$. The field equations are therefore overdetermined if both the conditions (56) and (34) and the variational equations (60) are implemented. Thus, these conditions can be used to determine, or at least restrict, the unknown function $L_0$, which must be a solution of the four equations (60), (56), (34). Interestingly, we use a variational problem here to determine not only the form of the Euler-Lagrange equations, but also the form of the Lagrange function itself.

As for the functional form of $L_0$ , we assume that $L_0$ does not depend on the derivatives of $\rho$ and $S$ with respect to time $t$; such a dependency would change the basic structure of the evolution equations. Further, if $L_0$ depends on $S$ and its derivatives, the validity of the continuity equation requires, as a consequence of (60), a complicated dependence of $L_0$ on the momentum distribution of the considered system. The simplest and at the same time most general $L_0$ can only depend on $\rho$ and its derivatives with respect to $q_k$, $L_0 = L_0(\rho, \partial_k\rho, \partial_k\partial_l\rho)$. Then, we find from Eqs. (60), (56), (34) that $L_0$ must obey the condition

$$\rho\frac{\partial L_0}{\partial \rho} - \frac{\partial}{\partial q_k}\rho\frac{\partial L_0}{\partial(\partial_k \rho)} + \frac{\partial}{\partial q_k}\frac{\partial}{\partial q_l}\rho\frac{\partial L_0}{\partial(\partial_k\partial_l \rho)} = 0. \tag{61}$$

The second order derivatives in $L_0$ must be taken into account but should not lead to contributions of higher order in the variational equations. The simplest nontrivial solution of (61) that satisfies this requirement is given by

$$L_0 = B_0\left[-\frac{1}{2\rho^2}\sum_k (\partial_k\rho)^2 + \frac{1}{\rho}\sum_k \partial_k\partial_k\rho\right], \tag{62}$$

where $B_0$ is a constant. The second term in the square bracket, multiplied by $\rho$, is a "Null-Lagrangian", which means that it does not contribute anything to the Euler- Lagrange equations [see (61)]. It is easy to verify that this $L_0$ also fulfills condition (58). If the variable $\rho$ is replaced by the amplitude $\sqrt{\rho}$ and the constant $B_0$ is replaced by $\hbar^2/4m$, one obtains the simpler expression

$$L_0 = \frac{\hbar^2}{2m}\frac{1}{\sqrt{\rho}}(\partial_k\partial_k\sqrt{\rho}). \tag{63}$$

Thus, the term $L_0$ agrees with the term $T_Q$ defined earlier, and the randomized field equation (56) agrees exactly with the modified Hamilton-Jacobi equation (47) obtained by linearization. This means that the process of linearization according to section 7 can be replaced by implementing the above statistical conditions, in particular Eqs. (51) and (59). The second transition, using statistical conditions, is slightly more involved, but enables a better understanding of the structure of QT. We will show in IV that this twofold path from QA to QT persists even when the full momentum field is taken into account.

### 8.1 Relation to Fisher information

If the term $\rho L_0$ is integrated over the configuration space, the second term in the square brackets of (62) gives no contribution and one obtains

$$\int \mathrm{d}q\, \rho\, L_0 = -\frac{B_0}{2}\, I[\rho], \tag{64}$$

where $I[\rho]$ is the Fisher information functional [18], defined by

$$I[\rho] = \int \mathrm{d}q\, \rho \sum_k \left(\frac{\partial_k \rho}{\rho}\right)^2. \tag{65}$$

While the second term in the square brackets of (62) gives no contribution to the variation it is nevertheless indispensable, because without it condition (58) could not be fulfilled. Fortunately, the same term is generated again when the first term of (62), i.e. the Fisher functional, is varied. The same final result is therefore obtained by adding the integrand of the Fisher functional to the Lagrangian density $\rho\bar{L}$ of the original system and performing the variation. In this way, Schrödinger's equation has already been derived by Reginatto using the given form (65) of the Fisher functional [52], see also [43].

The importance of the concept of Fisher information for the construction of the field equations of physics has been particularly emphasized by Frieden [18, 19]. The Fisher functional (65) takes its smallest value zero for spatially constant probability densities. Physically meaningful results are obtained if the requirement for minimal Fisher information is combined with suitable constraints. The situation is basically the same as with the entropy functional, which is written in configuration space as

$$S[\rho] = -\int \mathrm{d}q\, \rho(q) \ln \rho(q), \tag{66}$$

where a proportionality constant has been ommitted. If, as the simplest case, one considers discrete probabilities and only requires as an additional condition that the sum of all probabilities is 1, then one obtains the most fundamental result of probability theory from the requirement of maximum entropy, namely that all probabilities must be equal. Other constraints lead to the well-known probability distributions of statistical physics.

In contrast to the Boltzmann-Shannon entropy (66), the Fisher information (65) contains the first derivatives of the probability density. While varying $S[\rho]$ yields algebraic equations, varying $I[\rho]$ yields differential equations. One can view the Fisher information as a local variant of the global Boltzmann Shannon entropy. It defines what form the terms of a differential equation must have, so that passing from one point in the state space to the next, one is in accordance with the laws of probability theory.

This interpretation of the Fisher information is confirmed by a comparison with the "relative entropy" or Kullback-Leibler entropy [39], defined by

$$G[\rho, \chi] = -\int \mathrm{d}q\, \rho(q) \ln \frac{\rho(q)}{\chi(q)}, \tag{67}$$

where $\chi(q)$ is the initial, "prior", probability density relative to which $\rho$ is to be determined. We identify $\chi(q)$ with the probability distribution $\rho(q_1, ..., q_k + \Delta q, ..., q_n)$ obtained from $\rho$ by shifting the $k$-th coordinate $q_k$ by an amount $\Delta q$, and denote the resulting relative entropy by $\Delta G_k[\rho]$. For small $\Delta q$ one obtains in leading order the result

$$\Delta G_k[\rho] = -\frac{1}{2}(\Delta q)^2 \int \mathrm{d}q\, \rho(q) \left(\frac{\partial_k \rho}{\rho}\right)^2, \tag{68}$$

Thus, the Fisher information (65) is obtained, apart from a proportionality constant, by adding up the contributions $\Delta G_k[\rho]/(\Delta q)^2$ from all coordinates.

## 9 Poincaré invariant and quantization condition

The physically meaningful solutions of the Schrödinger equation must meet two additional conditions. First, the total probability of finding the particle(s) anywhere must be equal to one. This condition is also present in the classical theory from which QT was derived and needs no further explanation. Naturally, its influence on the form of the solutions is relatively strong for confined systems, that are located in a bounded spatial area. Second, the quantum mechanical state variable $\psi$, as defined by (39), must be a single-valued function in configuration space (at each instant of time). This allows for multi-valued functions $S/\hbar$ whose values differ from each other by multiples of $2\pi$. This single-valuedness condition is especially important for the determination of the spectra of atomic systems and was called "quantization condition" in the old quantum theory. The question arises whether or not an analogous structure exists in the field of classical physics. A candidate for an analogue structure is the concept of integral variants.

### 9.1 The general concept of integral invariants

In order to briefly recapitulate the concept of integral invariants, it is useful to consider non-autonomous dynamical systems, which are defined by the basic equations

$$\dot{x_k} = V_k(x, t), \tag{69}$$

where $V(x,t)$ is referred to as velocity field. Here, $x = (x_1, ..., x_m)$ denotes a point in a $m$-dimensional space $\mathbb{A}$. Let us consider an arbitrary closed contour $C_0$ in $\mathbb{A}$ at a fixed time $t_0$. The solutions of (69) transform each point of the contour $C_0$ to a different point at a later (fixed) time $t$. The totality of all transformed points forms again a closed contour $C_t$, at time $t$, as a consequence of continuity. The contour may change its form but remains a contour if transformed according to the flow; it is said that the contour is "moving with the fluid". The solutions of (69) may be written as $x_k(t,u)$ where $u$ is a parameter varying, say, in the interval $[0,1]$. For fixed $t$ the points $x_k(t,u)$ trace out a closed contour (with $x_k(t,0) = x_k(t,1)$), for fixed $u$ the points $x_k(t,u)$ describe trajectories. The totality of all trajectories is referred to as tube.

Consider now a vector field $G(x,t)$ on $\mathbb{A}$, which may also parametrically depend on time $t$. The contour integral

$$I(t) = \oint_{C_t} G_i(x,t)\mathrm{d}x_i \tag{70}$$

depends for given $C_0$ on the time $t$ and on the analytical form of the field $G(x,t)$. Evaluating the derivative of (70) with respect to time, one obtains the following necessary and sufficient condition for the invariance of $I(t)$ along the considered tube [48]:

$$\oint_{C_t} \mathrm{d}x_i \left\{ \frac{\partial G_i}{\partial t} + \left( \frac{\partial G_i}{\partial x_k} - \frac{\partial G_k}{\partial x_i} \right) V_k \right\} = 0. \tag{71}$$

If $G(x,t)$ fulfills condition (71) the contour integral (70) is referred to as a (linear, relative) "integral invariant".

## 9.2 Integral invariants in phase space

Let us now specialize this formalism to autonomous systems of Hamiltonian type. We have $\mathbb{A} = \Omega$, $m = 2n$, $x = (q_1,..,q_n,p_1,..,p_n) = (q,p)$ and

$$V(q,p,t) = \left( \frac{\partial H(q,p)}{\partial p}, -\frac{\partial H(q,p)}{\partial q} \right), \tag{72}$$

which means that (69) agrees with the canonical equations. The latter may also be written in the form $\dot{x} = Z\frac{\partial H}{\partial x}$ where $Z$ is an antisymmetric $2n \times 2n$ matrix with constant elements (the matrix $Z$ is referred to as symplectic matrix). It can be shown that essentially only a single integral invariant of type (70) exists, which is defined by $G(q,p,t) = (p_1,..,p_n,0,..,0)$. The corresponding line integral

$$\oint_{C_t} p_i \mathrm{d}q_i, \tag{73}$$

whose invariance is easily verified, is referred to as Poincaré integral invariant. The antisymmetric matrix in (71), which is the $2n-$dimensional counterpart of the rotation of a vector in three dimensions, agrees with the constant matrix $Z$

$$\frac{\partial G_i}{\partial x_k} - \frac{\partial G_k}{\partial x_i} = Z_{ik}. \tag{74}$$

Thus, the "vorticity tensor" of the vector field $G(q,p,t)$ is the same for all points of phase space, as a consequence of its simple (linear in $q$ and $p$) structure. The Poincare invariant cannot be used to characterize individual systems, but rather describes certain structural properties of the theory; the same applies to the conceptually similar but more general Poincare-Cartan invariant, which may be used as an axiomatic basis for the whole theory [20].

## 9.3 Transition to configuration space

This situation changes if the replacement $p_k \to M_k(q,t)$ is performed. After projection to the $n$-dimensional subspace defined by $M$, the Poincaré integral invariant takes the form

$$I_{\bar{C}}(t) = \oint_{\bar{C}_t} M_i(q,t)\mathrm{d}q_i. \tag{75}$$

Comparing the canonical condition (21) with the appropriate version of the invariance condition (71) one sees immediately that the latter is fulfilled if $h(q,t)$ is a single-valued function. In phase space arbitrary closed paths were allowed. Now, the allowed closed paths, denoted here by $\bar{C}_t$, as well as the tube formed by the solutions of (20), must lie in the subset $(q, M(q,t))$ of $\Omega$. As a consequence, the Poincaré invariant may now be used to characterize individual systems, it depends not only on the considered path but also on the initial value $M_0(q)$. The vorticity tensor $\Omega_{ik}(q,t)$, as defined by (30), is not a constant anymore but varies in configuration space in a way depending on the considered momentum surface. This means that vorticity for individual physical systems *does not exist in phase space* but only arises through the projection onto the configuration space. This is remarkable in view of the fact that vorticity is the (quasi-)classical [more precisely: the quasi-quantal] counterpart of quantum spin, as will be shown in IV.

The tensor $\Omega_{ik}(q,t)$ is the $n-$dimensional generalization of the fluid-dynamical "vorticity",

which is given by $\vec{\nabla} \times \vec{u}$ where $\vec{u}$ is the velocity field. The constant $I_{\bar{C}}$ is also referred to as "circulation". For $n = 3$ several of the above relations agree - after a proper reinterpretation of the variables - with the dynamic equations for an ideal barotropic fluid. In this fluid-dynamical context the invariance of the circulation is referred to as Kelvin's theorem.

In the next step we specialize to irrotational momentum fields (31). That means we enter the area of QA, where the Clebsch potential S becomes the quantum mechanical phase in the next step. In this case the circulation (75) vanishes for arbitrary $\bar{C}$ given that $S(q,t)$ is single-valued. The invariance condition (71) takes the form

$$\oint_{\bar{C}_t} \mathrm{d}q_i \frac{\partial h(q,t)}{\partial q_k} = 0, \tag{76}$$

if the derivative of $M$ with respect to time is replaced with the help of the Hamilton-Jacobi equation (33). This condition, the single-valuedness of $h(q,t)$, is the same as for general momentum fields. Vorticity and circulation vanish by definition, except in the presence of topological singularities of $S(q,t)$. In the present formalism QA there are, however, no natural causes for such topological singularities. We may say that Kelvin's theorem holds true in the QA (We remind the reader at this point that the theory QA is generally unstable, as discussed in section 6; we are considering a possibly small period of time in which stability prevails). The invariance of the circulation of S has a dynamic origin in the QA.

### 9.4 Transition to quantum theory

If we carry out the transition to QT as a last step, we obtain an unchanged continuity equation (34), and a modified Hamilton-Jacobi equation (47). These two differential equations are to be solved taking the above multi-valuedness condition for $S$ into account; if required by the considered physical system. The particle-like Eq. (32) played no role in the transition to QT; according to our discussion in section 7 this equation has no independent physical meaning in QT anymore. The validity of Kelvin's theorem in QA, as expressed by the validity of Eq. (76), is an indication of the fact that the concept of particle motion is still valid in QA. We can test Eq. (32) in this regard by seeing whether or not a conservation law of the type of Kelvin's theorem still exists in the QT. Using (47) the invariance condition (71) takes the form

$$\oint_{\bar{C}_t} \mathrm{d}q_i \frac{\partial}{\partial q_i} \left[ h(q,t) - \frac{\hbar^2}{2m} \frac{1}{\sqrt{\rho}} (\partial_k \partial_k \sqrt{\rho}) \right] = 0. \tag{77}$$

At first glance, this equation looks similar to (76), since the integrand is given by a gradient in both cases. Due to the coupling between the phase $S$ and the probability density $\rho$ in Eq. (47), however, there is now a natural source for singularities, since $S$ is undefined at the zeros of $\rho$ and may become discontinuous there. This happens quite often in QT and leads to a multivaluedness of $S$, which is characterized by an integer $n$ different from zero. The question is if this $n$ remains the same if the path $\bar{C}_t$ changes as a function of time according to the differential equation (32). Damski and Sacha have shown that this is generally not the case by examining several simple physical systems [11]. Vortices in the quantum mechanical probability fluid are rather unstable. Vortices in many-particle systems (superfluid phases), which are described by non-linear field equations, are more stable.

This means that Kelvin's theorem does not hold true in the QT. There is no support, from this point of view, for the idea that Eq. (32) describes the motion of real particles. The analogy between Kelvin's theorem and the quantization condition is of a formal nature. Kelvin's theorem is of dynamical origin, while the quantization condition of QT has a topological cause, namely the uniqueness of the state variable $\psi$. In the present framework of HLLK the latter was introduced due to the requirement for *linearity* of the basic equations.

In the above derivation we have assumed an irrotational momentum field $M$. This means that we have made the unphysical assumption of strong functional dependencies between the components $M_k$. In the future work IV the above derivation will be generalised to arbitrary functionally independent, i.e. realistic momentum fields, which also describe the phenomenon of spin.

# 10 Which objects can be described by QT?

We now return to the question we posed at the beginning of this paper. The reconstruction of QT on the basis of PM that we have carried out in I, II, and the present work represents a success of the ensemble interpretation. We built QT by systematically going through the transitions between theories: type 1 ⇒ type 2 ⇒ type 3. One must not jump from type 1 to type 3, as is done in the context of the individuality interpretation, but must take into account the statistical uncertainty already in the classical area. By choosing type 2 as starting point, the object that the theory covers becomes a statistical ensemble. The transition to type 3 does not change this. This takes a fundamentally indeterministic standpoint when explaining the microcosm [50].

Each of the steps we have taken results in a *loss of information*. This becomes particularly clear if the step QA ⇒ QT is carried out using randomization instead of linearization. This means that neither the PM nor the CM can be derived from QT. However, such a reduction of QT to classical physics, in the limit $\hbar \to 0$, is central to the widespread opinion that QT can describe individual particles, as noted already in the introduction. Let us analyze this question in more detail.

## 10.1 Formal versus empirical reduction

What exactly does it mean when we say that the better theory (QT) is reduced to the worse (CM or PM)? In order to answer this question we have to fall back on some very elementary facts. A physical theory is defined by a certain number of fundamental *equations*, mostly differential equations. With the help of the solutions of the fundamental equations (of a successful theory) one can describe a certain range of processes in nature. It is the *solutions* that decide whether a theory is successful or not.

The first clearly defined meaning of the term reduction is obtained when one considers the relationship between the fundamental *equations*. This means, for example, that the QT is reduced (in this sense) to the CM or to the PM if Schrödinger's equation in the limit of small $\hbar$ merges into the canonical equations or into the Liouville equation. An important point is, that this kind of reduction has nothing to do with the question of the truth or falsity of the theories in question. For example, if the Liouville equation could be reduced to Schrödinger's equation, that would not mean that the Liouville equation is the better theory. Following Rosaler [53], we refer to this kind of reduction as *formal reduction*.

It is unsatisfactory to characterize the relationship between two theories solely on the basis of the formal relationship between their fundamental equations; the question which one of the two theories describes nature correctly, is of course of paramount importance and must also be taken into account. In order to introduce a meaningful concept of reduction from this point of view, one must compare the *solutions* of the corresponding theories - and the associated experimental results - with one another: The successful theory A is reduced to theory B, in a certain area, if the results of B are reproduced at least approximately in this area. Clearly, the concept of reduction has a slightly different meaning here and, due to the large number of possible solutions (systems), is by far not as sharply defined as in the formal case. Using again the notation of Rosaler [53] we refer to this kind of reduction as *empirical reduction*.

## 10.2 The quantum-classical interface

The relationship between QT and classical physics has not been analytically investigated for an astonishingly long period of time; one can say that the majority approach was dominated by a philosophical principle, the principle of reductionism. In particular, no distinction was made between formal reduction and empirical reduction.

Let us start with the question of formal reduction, which is the one studied in the present work. The first systematic investigation of this question [37] led to the conclusion that, contrary to common belief, QT is not formally reducible to CM. A second recent study, reported in II, which took into account all the theoretical results obtained during the past decades, led to the stronger conclusion that QT can neither be reduced to CM nor can it be reduced to the prob-abilistic version PM of classical mechanics. In II it was shown that, to the contrary, PM may be reduced to QT. In the present work we arrived at the same conclusion, using a different method that allows us to study the details of the transition. This represents a radical change as compared to the common view. The projection method used in the present work can still be called a reduction, since phase space is larger than configuration space. However, this type of formal reduction is not associated with a vanishing $\hbar$, but to the contrary with the *creation* of this new fundamental constant. Its finite numerical value is a consequence of the randomization, or linearization, following the projection.

Let us proceed now to the question of empirical reduction. There are several systems in which an empirical reduction of QT to classical physics can be observed. The historically oldest case concerns transitions between highly excited atomic states. Here, an analogy to classical oscillators exists which is described by Bohr's correspondence principle. If the characteristic dimensions of a system are much larger than the De Broglie wavelength, then a classical approximation is often allowed. Generally speaking, the influence of typical quantum mechanical effects decreases when a characteristic quantity of the dimension of an action can be defined, whose numerical value is much larger than $\hbar$. In more complex systems, high particle numbers and high temperature have a similar effect. While all these effects have only an approximate character, they are undoubtedly real, so it is justified to speak of an empirical reduction of QT to classical physics.

Thus, there are two interfaces between QT and classical physics, a formal and an empirical one. Both interfaces are associated in a natural way with a direction. The formal interface is associated with the direction from classical physics (PM) to QT, the empirical interface is associated with the direction from QT to classical physics. In both cases we have transitions from a more complicated structure to a simpler one, that is why we can speak of a reduction in both cases. For the formal reduction, the direction points from the phase space to the lower dimensional configuration space. For the empirical reduction the direction points from QT to (quasi-) classical systems, in which the fine structure of the complicated quantum effects is "smeared out".

## 10.3 Pauli's mechanical paths and Dirac's families of solutions

Let us return to the question whether QT is able to describe the behavior of individual particles or whether sensible statements can only be made about statistical ensembles. The studies carried out here and in I, II and in earlier works [37] lead to the conclusion that QT is only suitable for describing statistical ensembles. This is by no means a new discovery. For example, Schrödinger already arrived at the same conclusion when analyzing the equation he had derived:

> "... the true laws of quantum mechanics do not consist of definite rules for the single path ..." [60].

Two comments should be made here for clarification. First, the above conclusion does not mean that individual particles cannot *exist*; we just have no theory to describe them (and do not believe that such a theory exists). Second, it is well known that the probabilistic structure of QT cannot be captured by standard (Kolmogorov) probability theory. However, this does not change the fundamental conclusion.

This conclusion is at variance with the reductionistic view, that the classical limit of QT must be CM. One can find in the literature several formal derivations that aim to prove the reductionist point of view. All of these derivations must be flawed somewhere if the above claim is to be correct. And, as expected, the corresponding fallacies are not easy to detect. For exampe, in section 31 of his famous book on quantum mechanics [14], Dirac claims that all quantum mechanical wave packets in the limiting case $\hbar \Rightarrow 0$ become sharp distributions with peaks at the classical orbits. This catchy statement has undoubtedly found its way into many classrooms. Unfortunately, this statement only applies to a small number of selected wave packets; the details of the analysis may be found in an earlier paper by the present author [37]. Two more of these subtle fallacies are discussed below.

If one examines the behavior of Schrödinger's equation for small $\hbar$, one obtains the Hamilton-Jacobi equation and the continuity equation as leading terms of an asymptotic expansion. These differential equations agree with Eqs. (33) and (34), which were obtained by projection and subsequent introduction of potentials. On the other hand, the Hamilton-Jacobi equation also occurs in the theory of canonical transformations [62]. If a complete solution $F_1(q_1,..,q_n,a_1,..,a_n)$ , which depends on $n$ parameters $a_1,..,a_n$, is known, the trajectories of all particles can be calculated. It was believed that this fact marked the transition from QT to CM. A complete solution with fixed numerical values of the parameters $a_1,..,a_n$ was interpreted as (quasi-)classical counterpart of the quantum mechanical state vector. Pauli wrote in his "General principles of quantum mechanics" ( [49], chapter VI, p. 91, bracket mine):

> If in the solution $S_0$ of ...[the Hamilton-Jacobi equation] no further parameter is present, it is one mechanical path and in the general case giving special numerical values to parameters $a_1,..,a_n$, ...occurring in $S_0$ corresponds to a definite trajectory".

In fact, a fixed choice of the parameters $a_1,..,a_n$ does not correspond to a single trajectory but to an $n$-fold infinite set - a "family" - of trajectories. This leads to a contradiction to the idea that the wave function describes a single particle. Dirac expressed this contradiction in the following frequently quoted statement [13]:

> The family does not have any importance from the point of view of Newtonian mechanics; but it corresponds to one state of motion in the quantum theory, so presumably the family has some deep significance in nature, not yet properly understood.

This contradiction persists to this day as one of the many contradictions that exist in the individuality interpretation of QT [33].

This line of argument is flawed as a consequence of the fact that the solutions of Schrödinger's equation are determined by their initial values. Therefore, the Hamilton-Jacobi equation (33), which is obtained for small $\hbar$, has also to be solved as an initial value problem and not as a complete solution problem. As a consequence, it is not immediately possible to set the quasi-classical limit of Schrödinger's equation in correspondence to CM. In order to make it possible, Pauli and Dirac considered special solutions, which are obtained by assigning fixed numerical values to the $n$ parameters $a_1,..,a_n$ of a complete solution, as solutions to the initial value problem. However, this is not mathematically consistent. As is known from the theory of partial differential equations, one can construct arbitrary solutions to initial value problems if a complete solution is known [17, 41]. However, this requires the calculation of the envelope of a whole family of solutions with different values of the parameters $a_1,..,a_n$; it is not possible to describe arbitrary initial conditions by simply assigning numerical values to the parameters. This is the error in the assertion that CM is the limiting case of QT for small $\hbar$. Let us mention once again that Eqs. (33) and (34) have to be solved as initial value problems. Thus, the quasi-classical limit of Schrödinger's equation fully agrees with the present theory QA.

As far as Dirac's statement is concerned, it is in the context of the ensemble interpretation obviously true (by construction) that a whole ensemble of classical trajectories corresponds to a single quantum state. Dirac's ensemble has no special significance in the present theory, since here the ensembles are defined by initial conditions and not by complete solutions; the "deep significance" may of course be seen in the general validity of this correspondence. The ensemble interpretation is seldom able to answer the questions posed by the individuality interpretation and vice versa [42]. Unfortunately, there is no common basis for formulating meaningful questions.

## 11 Discussion

This series of works is based on the fact that there are mathematical structures in phase space that are very similar to mathematical structures of QT. This similarity becomes fully visible only when statistical ensembles of particles, instead of single particles, are considered.

Many of these structures are described in Sudarshan and Mukunda's outstanding book on analytical mechanics [62]. The solutions of the canonical equations for any observable $A(q,p)$, which takes the place of Hamilton's function, form a one-parameter group of canonical transformations in phase space (a realization of a corresponding one-parameter Lie group). These transformations are the classical counterparts of the unitary transformations in Hilbert space; the observables $A(q,p)$ correspond to the self-adjoint operators that are the generators of these unitary transformations. The vanishing of the Poisson bracket of two observables $A$ and $B$ implies mutual invariance under the transformations generated by the other observable. As is well-known, the Poisson bracket of two observables is the classical counterpart of the commutator of the corresponding operators in QT. Each observable $A$ creates a statistical ensemble with the help of a velocity field $V_A$, a probability density $\rho_A$, and a corresponding Liouville equation. In this work we have identified $A$ with the Hamiltonian function $H$. The present method may, however, be used for any observable and leads to the same results as in II.

This summary shows that the mathematical and physical analogy between functions in phase space (describing states of ensembles) and vectors in Hilbert space is much stronger than the corresponding analogy for points in phase space (describing states of particles). Accordingly, only a few simple and natural additional assumptions have to be made, in the framework of HLLK, in order to derive QT from the above classical probabilistic structure PM. The first assumption is the introduction of a new dynamical variable $S(q,p,t)$ that describes the purely deterministic content of the canonical equation in a compact way. A second fundamental assumption is that a projection to configuration space must be performed. This is plausible from a physical point of view; it is downright trivial if one considers the simplest formulation of QT, on the level of differential equations, and looks at the independent variables. The final result of the HLLK is a construction of QT that provides not only the basic differential equation, but also most of the other structural elements of QT, such as the probability interpretation of the wave function, the explanation of why observables become operators, the eigenvalue equations for measurable values of operators, and Born's rule. A more complete list of structural elements of QT may be found in II. In the following work IV this list will be completed and to a certain extent closed, by showing that the HLLK also provides an explanation for the spin of QT. Anticipating this point, we may say that QT as a whole "emerges" from PM, if the assumptions of the HLLK are implemented. Of course, if the transition from CM to PM is included as first step of the theory, we may also say that QT emerges from CM. The HLLK brings together numerous known results and some new considerations under a common idea, which was essentially formulated by Einstein. It provides a new conceptual foundation for QT that seems to be simpler, and at the same time more complete, than any other attempt in this direction.

## 12 Concluding remarks

The HLLK is based on the assumption that QT can be derived from classical probabilistic mechanics by means of a projection and a linearization. This concept has been confirmed in the present work; we have performed the same steps as in I and II, just in different order. This required a different mathematical realization of the fundamental concept of the HLLK but led to the same final result. So far we have only considered the non-relativistic description of massive particles without spin. One may wonder whether the basic concept of the HLLK is applicable to other areas. This is indeed the case. We have already mentioned that the complete description of the momentum field, that is, the consideration of all three Clebsch potentials, leads to the phenomenon of spin. If one takes into account the relativistic space-time structure in the particle equations of motion, one can derive the Dirac equation, as will be shown in a later paper. Remarkably, it is even possible to apply the method to massless particles. The spin 1 wave equation obtained with a variant of the method used in I, II is equivalent to Maxwell's equations in empty space [34]. In this respect, the HLLK appears to play an expanded role, namely as a method for deriving fundamental field equations. The question arises whether it can be used to derive *all* fundamental field equations, and only these.

## Acknowledgments

The author thanks Daniel Rohe for stimulating e-mail correspondence.

## References

[1] V. I. Arnold. *Ordinary Differential Equations.* Berlin: Springer, 1992.

[2] L. E. Ballentine. “The statistical interpretation of quantum mechanics”. In: *Reviews of Modern Physics* 42 (1970), pp. 358–381.

[3] A. Bennett. *Lagrangian fluid dynamics.* Cambridge, UK: Cambridge University Press, 2006.

[4] M. V. Berry and N. L. Balazs. “Evolution of semiclassical quantum states in phase space”. In: *J. Phys. A* 12 (1979), pp. 625–642.

[5] A. O. Bolivar. “The Wigner representation of classical mechanics, quantization and classical limit”. In: *Physica* A 301 (2001), pp. 219–240.

[6] F. Bopp. “The principles of the statistical equations of motion on quantum theory”. In: *Observation and Interpretation.* Ed. by S. Körner. London: Butterworths Scientific Publications, 1957, p. 189.

[7] R. Bopp. “La mecanique quantique est-elle une mecanique statistique classique particuliere ?” In: *Ann. Inst H. Poincare* 15 (1956), pp. 81–112.

[8] M. Born. “Vorhersagbarkeit in der Klassischen Mechanik”. In: *Zeitschrift für Physik* 153 (1958), pp. 372–388.

[9] C. Caratheodory. *Calculus of Variations and Partial Differential Equations of the First Order, Part I.* San Francisco: Holden-Day, Inc, 1965.

[10] A. Clebsch. “Über eine allgemeine Transformation der hydrodynamischen Gleichungen”. In: *J. reine angew. Math.* 54 (1857), pp. 293–312.

[11] B. Damski and K. Sacha. “Changes of the topological charge of vortices”. In: *J. Phys. A: Math. Gen.* 36 (2003). see also: https://arxiv.org/abs/quant-ph/0202137v1, pp. 2339–2345.

[12] J. B. Delos. “Catastrophes and stable caustics in bound states of Hamiltonian systems”. In: *J. Chem. Phys.* 86 (1987), pp. 425–439.

[13] P. A. M. Dirac. “The Hamiltonian form of field dynamics”. In: *Can. J. Math.* 3 (1951), pp. 1–23.

[14] P. A. M. Dirac. *The principles of quantum mechanics.* 4. p. 88. Oxford: Oxford University Press, 1958.

[15] A. Einstein. “Reply to Criticism”. In: *Albert Einstein: Philosopher-Scientist.* New York: Harper and Row, 1949, p. 665.

[16] A. Einstein, B. Podolsky, and N. Rosen. “Can quantum-mechanical description of physical reality be considered complete?” In: *Phys. Rev.* 47.10 (May 1935), pp. 777–780.

[17] L. C. Evans. *Partial differential equations.* Providence: American Mathematical Society, 1998.

[18] B. Roy Frieden. *Physics from Fisher Information, a Unification.* Cambridge: Cambridge University Press, 1998.

[19] B. Roy Frieden and B. Soffer. “Lagrangians of physics and the game of Fisher-information transfer”. In: *Phys. Rev. E* 52 (1995), p. 2274.

[20] F. Gantmacher. *Lectures in Analytical Mechanics.* Moscow: Mir Publishers, 1975.

[21] H. J. Groenewold. “On the principles of elementary quantum mechanics”. In: *Physica* 12 (1946), pp. 405–460.

[22] B. C. Hall and M. Reginatto. *Ensembles on Configuration Space.* New York, Heidelberg, Dordrecht: Springer International Publishing, 2016.

[23] M. J. Hall and M. Reginatto. “Schrödinger equation from an exact uncertainty principle”. In: *J. Phys. A* 35 (2002), pp. 3289–3303.

[24] T. Hawkins. “Frobenius, Cartan and the Problem of Pfaff”. In: *Arch. Hist. Exact. Sci.* 59 (2005), pp. 381–436.

[25] S. Hayakawa. “Atomism and Cosmology”. In: *Suppl. Progr. Theor. Phys.* 34 (1965), pp. 533–541.

[26] P. R. Holland. *The quantum theory of motion.* Cambridge, U.K.: Cambridge University Press, 1995.

[27] C. Jaffe and P. Brumer. “Classical Liouville Mechanics and Intramolekular Relaxation Dynamics”. In: *J. Phys. Chem.* 88 (1984), pp. 4829–4839.

[28] A. Khrennikov. “Bell-Boole inequality: Nonlocality or Probabilistic Incompatibility of Random Variables?” In: *Entropy* 10 (2008), pp. 19–32.

[29] U. Klein. “A reconstruction of quantum theory for spinning particles”. arXiv:2202.13364.

[30] U. Klein. “A statistical derivation of non-relativistic quantum theory”. In: *Measurements in Quantum Mechanics*. Ed. by M. R. Pahlavani. see also arXiv:1109.6244 [quant-ph]. Rijeka, Croatia: ISBN:978-953-51-0058-4, 2012, pp. 141–174.

[31] U. Klein. “From Koopman-von Neumann theory to quantum theory”. In: *Quantum Stud.: Math. Found.* 5 (2018), pp. 219–227.

[32] U. Klein. “From probabilistic mechanics to quantum theory”. In: *Quantum Stud.: Math. Found.* 7 (2020), pp. 77–98.

[33] U. Klein. “Is the individuality interpretation of quantum theory wrong ?” arXiv:1207.6215 [quant-ph].

[34] U. Klein. “Quantizing Galilean spacetime: a reconstruction of Maxwell’s equations in empty space”. In: *Quantum Stud.: Math. Found.* 11 (2024), pp. 717–737.

[35] U. Klein. “Schrödinger’s equation with gauge coupling derived from a continuity equation”. In: *Foundations of Physics* 39 (2009), p. 964.

[36] U. Klein. “The statistical origins of quantum mechanics”. In: *Physics Research International* 2010 (). http://downloads.hindawi.com/journals/phys/2010/808424.pdf, p. 18.

[37] U. Klein. “What is the limit $\hbar \to 0$ of quantum theory?” In: *Am. J. Phys.* 80 (2012). see also http://arxiv.org/abs/1201.0150, p. 1009.

[38] V. V. Kozlov. “General Theory of Vortices”. In: *Dynamical Systems X*. Vol. 67, Encyclopaedia of Mathematical Sciences. Berlin: Springer, 2003.

[39] S. Kullback. *Information Theory and Statistics*. New-York: Wiley, 1959.

[40] M. Kupczynski. “Is quantum theory predictably complete?” In: *Physica Scripta* T135 (2009), p. 014005.

[41] C. Lanczos. *The variational principles of mechanics*. Toronto: University of Toronto Press, 1952.

[42] N. P. Landsman. “Between classical and quantum”. In: *Handbook of the Philosophy of Physics, Vol. 2*. Ed. by J. Butterfield and J. Earman. available at: http://arxiv.org/abs/quant-ph/0506082. Amsterdam: Elsevier, 2005.

[43] Y. C. Lee and W. Zhu. “The principle of minimal quantum fluctuations for the time-dependent Schrödinger equation”. In: *J. Phys. A* 32 (1999), pp. 3127–3131.

[44] E. Madelung. “Quantentheorie in hydrodynamischer Form”. In: *Z. Phys.* 40 (1926), pp. 322–326.

[45] J. E. Moyal. “Quantum Mechanics as a Statistical Theory”. In: *Proc. Cambridge Phil. Soc.* 45 (1949), p. 99.

[46] N. Mukunda. “Phase space methods and the Hamilton-Jacobi form of dynamics”. In: *Proc. Indian Acad. Sci.* 87 (1987), pp. 85–105.

[47] L. S. F. Olavo. “Foundations of quantum mechanics: non-relativistic theory”. In: *Physica* A 262 (1999), pp. 197–214.

[48] L. A. Pars. *A Treatise on Analytical Dynamics*. London: Heinemann, 1965.

[49] Wolfgang Pauli. *General principles of quantum mechanics*. chapter VI. Berlin: Springer, 1980, p. 212.

[50] Karl R. Popper. *The Open Universe - an Argument for Indeterminism*. Totowa: Rowman and Littlefield, 1982.

[51] H. De Raedt, M. I. Katnelson, and K. Michielsen. “Quantum theory as the most robust description of reproducible experiments”. In: *Ann. Phys.* 347 (2014), pp. 45–73.

[52] M. Reginatto. “Derivation of the equations of nonrelativistic quantum mechanics using the principle of minimum Fisher information”. In: *Phys. Rev. A* 58 (1998), pp. 1775–1778.

[53] J. Rosaler. “Formal vs. Empirical Approaches to Quantum-Classical Reduction”. In: *Topoi* 34 (2015), pp. 325–338.

[54] N. Rosen. “The relation between classical and quantum mechanics”. In: *Am. J. Phys.* 32 (1964), pp. 597–600.

[55] P. C. Rosenbloom. “The Hamilton-Jacobi equation”. In: *Arch. Rational Mech. Anal.* 3 (1971), pp. 245–254.

[56] H. Rund. “Clebsch potentials and variational principles in the theory of dynamical systems”. In: *Arch. Ration. Mech. Anal.* 65 (1977), p. 305.

[57] H. Rund. “Clebsch representations and relativistic dynamical systems”. In: *Arch. Ration. Mech. Anal.* 71 (1979), p. 199.

[58] R. Schiller. “Quasi-Classical Theory of the Nonspinning Electron”. In: *Phys. Rev.* 125.3 (Feb. 1962), pp. 1100–1108.

[59] W. P. Schleich et al. “Schrödinger equation revisited”. In: *Proc. Natl. Acad. Sci. USA* 110.14 (2013), pp. 5374–5379.

[60] Erwin Schrödinger. “Quantisierung als Eigenwertproblem, Zweite Mitteilung”. In: *Annalen der Physik* 79 (1926), p. 489.

[61] L. Skala, J. Cizek, and V. Kapsa. “Quantum mechanics as applied mathematical statistics”. In: *Ann. Phys.* 326 (2011), pp. 1174–1188.

[62] E. C. G. Sudarshan and N. Mukunda. *Classical Dynamics: A Modern Perspective.* New York: Wiley, 1974.

[63] T. Takabayasi. “On the formulation of quantum theory associated with classical pictures”. In: *Progress in Theoretical Physics.* 8.2 (1952), pp. 143–182.

[64] K. Takahashi. “Distribution functions in classical and quantum physics”. In: *Progress in Theoretical Physics Supplement.* 98 (1989), pp. 109–156.

[65] J. H. Van Vleck. “The correspondence principle in the statistical interpretation of quantum mechanics”. In: *Proc. Natl. Acad. Sci. U.S.* 14 (1928), pp. 178–188.

[66] T. C. Wallstrom. “Inequivalence between the Schrödinger equation and the Madelung hydrodynamic equations”. In: *Phys. Rev. A* 49.3 (1993), pp. 1613–1617.

[67] C. Wetterich. “Quantum particles from classical probabilities in phase space”. In: *Int. J. Theor. Phys.* 51 (2012), pp. 3236–3273.